\documentclass[%
 reprint, amsmath,amssymb,aps,
]{revtex4-2}

\usepackage{amsmath}
\usepackage{graphicx}
\usepackage{dcolumn}
\usepackage{bm}
\usepackage{booktabs}
\usepackage[capitalize]{cleveref}

\usepackage{graphicx}
\usepackage{caption}                                                           
\usepackage{subcaption}
\usepackage{comment}
\makeatletter
\DeclareRobustCommand{\varname}[1]{\begingroup\newmcodes@\mathit{#1}\endgroup}
\makeatother

\begin{document}

\preprint{APS/123-QED}

\title{Statistical causal inference methods for observational research in PER: a primer}

\author{Vidushi Adlakha}
\author{Eric Kuo}%
\affiliation{%
University of Illinois at Urbana-Champaign, Urbana, Illinois 61820, USA
}%




\date{\today}

\begin{abstract}
Recent critiques of Physics Education Research (PER) studies \cite{weissman2021policy,weissman2022invalid} have revoiced the critical issues when drawing causal inferences from observational data where no intervention is present.  In response to a call for a ``causal reasoning primer" in PER \cite{weissman2021policy}, this paper discusses some of the fundamental issues underlying statistical causal inference. In reviewing these issues, we discuss well-established causal inference methods commonly applied in other fields and discuss their application to PER.  Using simulated data sets, we illustrate (i) why analysis for causal inference should control for confounders but not control for mediators and colliders and (ii) that multiple proposed causal models can fit a highly correlated data set.  Finally, we discuss how these causal inference methods can be used to represent and explain existing issues in quantitative PER.  Throughout, we discuss a central issue: quantitative results from observational studies cannot support a researcher's proposed causal model over other alternative models.  To address this issue, we propose an explicit role for observational studies in PER that draw statistical causal inferences: proposing future intervention studies and predicting their outcomes.  Mirroring a broader connection between theoretical motivating experiments in physics, observational studies in PER can make quantitative predictions of the causal effects of interventions, and future intervention studies can test those predictions directly.
 
\end{abstract}

\maketitle


\section{Introduction}

Recent critiques of Physics Education Research (PER) studies \cite{weissman2021policy,weissman2022invalid} have revoiced the critical issues when drawing causal inferences from observational data where no intervention is present. In response to a call for a ``causal reasoning primer" in PER \cite{weissman2021policy}, this paper discusses some of the fundamental issues underlying statistical causal inference. In reviewing these issues, we discuss well-established causal inference methods commonly applied in other fields \cite{varian2016causal,rohrer2018thinking,foster2010causal,glass2013causal,gangl2010causal,keele2015statistics,murnane2010methods,vandenbroucke2016causality,gangl2010causal,freedmanbook,imbens2020potential,pearce2016causal,williams2018directed,imbens2015causal} and discuss their application to PER.  We suspect that many Physics Education researchers engaged in quantitative analysis will be familiar with these methods.  At the same time, we propose that more widespread knowledge of these causal inference methods can help establish greater consensus among the PER community on how to establish causal relationships from quantitative data. The causal inference methods we present provide a powerful set of conceptual and mathematical tools for analysis and make clear the potential causal misinterpretations and biases that can be introduced during analysis.  For readers interested in a more in-depth discussion of the causal methods discussed here, there are both more popular \cite{pearl2018book} and technical references \cite{greenland1999causal,hernan2020book,glymour2016causal,glymour2019review,morgan2013handbook,peters2017elements,morgan2015counterfactuals,pearl2013linear} available. 

\section{Causal vs. Predictive Modeling}
In this paper, we will use path analysis \cite{wright1921correlation,wright1918nature,wright1934method} using multiple linear regression on standardized variables to illustrate causal inference methods, though the causal issues we illustrate extend to other analytic methods as well (e.g., structural equation modeling \cite{saris1983introduction}). 

Consider the case where three standardized  variables, $X$, $Y$, and $Z$, are measured, and a multiple regression is performed with $Z$ as the dependent variable and $X$ and $Y$ are independent variables (an analysis denoted as $Z \sim X + Y$). This best-fit linear model produced by this analysis is $Z = \beta_{XZ} X + \beta_{YZ} Y$ (note: there will be a non-zero constant term $\beta_0$ if the variables are not all standardized). Conceptually, this analysis is commonly interpreted as finding the variance explained by one independent variable while controlling for another (i.e. finding the regression coefficient of $X$ on $Z$, $\beta_{XZ}$, when controlling for $Y$).  For this regression analysis, $\beta_{XZ}$ is:

\begin{equation} \label{eq:regcoeff}
\beta_{XZ} = \frac{r_{XZ}-r_{XY} r_{YZ}}{1 - r_{XY}^2}
\end{equation}

A conceptually important limiting case is that when $r_{XY}=0$, $\beta_{XZ}=r_{XZ}$.  This indicates that controlling for the independent variable $Y$ has no effect on the association between independent variable $X$ and dependent variable $Z$ if the two independent variables are not correlated.

The interpretation and appropriateness of the analysis $Z \sim X + Y$ will depend on whether the goal of this regression model is predictive or causal \cite{shmueli2010explain}.  For a predictive model \cite{kuhn2013applied}, the goal would be to explain the most variance in $Z$ with other measured variables - that is, to reduce the error in predicting $Z$.  One example would be using early pre- and in-course measures to predict students' final physics course grade \cite{burkholder2021importance,mccammon1988predicting,verostek2021analyzing,cwik2022students,salehi2019demographic,huang2013predicting,ting2001predicting,ransdell2001predicting,pokay1990predicting,cohen2017research,zabriskie2019using,sadler2001success,hazari2007gender,kitsantas2008self,rimfeld2016true}.  Establishing this predictive model using data from previous semesters may allow instructors and researchers to identify which students are at-risk of failing a course early enough to provide additional support.  In a predictive model, it is sensible to include as many variables as available to improve $R^2$ of the model - it does not matter what $X$ or $Y$ represents. The $\beta$'s indicate which variables explain the most variance in $Z$ that is not explained by other variables in the model - that is, the variance explained by one independent variable controlling for all others.

By contrast, the goal of a causal model \cite{pearl2018book} is to estimate the causal impact of how intervening on $X$ and $Y$ should affect $Z$.  That is, if $X$ is changed by one standard deviation, how would $Z$ change?  A common pitfall is assuming that $\beta_{XZ}$ is an accurate estimate of this causal impact when, in actuality, this depends on the proposed causal model of how $X$, $Y$, $Z$, and other unmeasured variables are related. To illustrate how the predictive and causal inference goals of statistical modeling can be misaligned, consider the case where $X$ is a student's mid-semester score in their math course.  In this case, although mid-semester calculus grade may help predict final physics course grade, interventions to improve mid-semester calculus grade may not improve final course grade.  For instance, an intervention that increases time spent on calculus study might actually reduce the time available for studying physics, causing no improvement or even decreasing students' physics final grade.  In reality, mid-semester calculus grade could serve as a proxy indicator of the causal role of students' more general math preparation or general study practices rather than on calculus-specific performance.  Although a predictive model does not necessarily care why variance in outcomes is explained, the causal details of why variance is explained are critical for making accurate causal estimates.  The rest of this paper elaborates on causal inference techniques for determining the appropriate analysis for estimating the causal impacts of one variable on another when many variables are correlated together. Central to these methods are diagrams that embody a theoretical model of the cause-effect relationships between variables.  

\section{Three Fundamental Causal Structures: Chain, Fork, Collider}

Relations between quantitative variables can be represented through directed acyclic graphs (DAGs) \cite{morgan2013handbook,pearl1995causal,spirtes2000causation}, which represent the variables as nodes connected by directed arrows.  DAG-like diagrams are commonly used to represent the results of path analysis or structural equation modeling.  When the DAG is constructed to reflect a proposed causal model, then the arrows indicate the direction of causality between variables, and coefficients associated with each arrow reflect the direct causal impact of changing one variable on another.  

For instance, $X \rightarrow Y$ is a causal model where $X$ has a causal impact on $Y$ – that is, intervening to change $X$ will produce a change in $Y$ and that intervening to change $Y$ directly (i.e., through a method besides changing $X$) will \emph{not} change $X$.  The analysis $Y \sim X$ would produce the coefficients of the linear equation $Y = \beta_{XY} X$, and the coefficient $\beta_XY$ would be associated with the path connecting $X$ to $Y$.  This equation (and diagram) also represents a quantitative causal prediction: that changing $X$ by $\Delta X$ will change $Y$ by $\beta_{XY}(\Delta X)$.  

There are three fundamental causal structures – chain, fork, and collider \cite{pearl2018book} – through which more complicated causal models can be constructed.  These three structures  illustrate the ways in which correlation may or may not reflect causation and also the different rules for how controlling for variables impacts analysis.  

\begin{figure}
     \centering
     \begin{subfigure}[b]{0.2\textwidth}
         \centering
         \includegraphics[width=\textwidth]{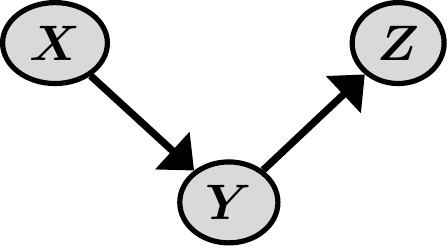}
         \caption{}
         \label{ChainDAG}
     \end{subfigure}
     \hfill
     \begin{subfigure}[b]{0.2\textwidth}
         \centering
         \includegraphics[width=\textwidth]{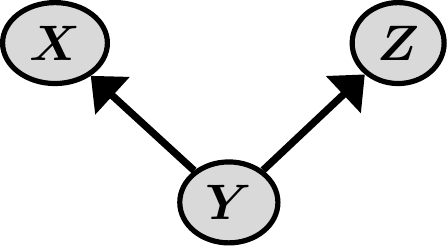}
         \caption{}
         \label{ForkDAG}
     \end{subfigure}
     \hfill
     \begin{subfigure}[b]{0.2\textwidth}
         \centering
         \includegraphics[width=\textwidth]{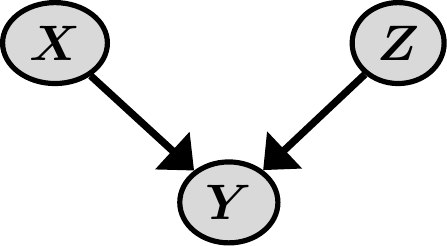}
         \caption{}
         \label{ColliderDAG}
     \end{subfigure}
        \caption{Fundamental causal structures: (a) Chain - where $Y$ acts as a mediator, (b) Fork - where $Y$ acts as a confounder, and (c) Collider - where $Y$ acts as a collider.}
        \label{fig:Causaldiagrams}
\end{figure}

\subsection{Chain}
A causal chain is represented as $X \rightarrow Y \rightarrow Z$ (Fig. \ref{ChainDAG}). This chain represents a causal mediation where $X$ causes $Z$ through the mediator $Y$: $X$ causes $Y$ and $Y$ causes $Z$, so therefore $X$ causes $Z$.  An everyday example of a causal chain is $Fire \rightarrow  Smoke \rightarrow Alarm$. Here, smoke is the mediator caused by fire and causes the smoke alarm to sound.  In principle, any single causal link can be modeled as a chain by explicitly breaking down the causal mechanism into mediators.  In practice, mediators are commonly omitted from causal diagrams if they are not measured and/or are not of theoretical interest.

The path coefficients of the chain $X \rightarrow Y \rightarrow Z$ are associated with two linear regressions, $Y = \beta_{XY}X$ and $Z = \beta_{YZ}Y$.  Because we have assumed $X$, $Y$, and $Z$ are standardized variables and the regressions only have a single independent variable,  $\beta_{XY} = r_{XY}$ and $\beta_{YZ} = r_{YZ}$.  There are two ways to determine the (indirect) causal impact of $X$ on $Z$.  The first is the chain rule: changes in $X$ cause changes in $Y$, and these changes in $Y$ cause changes in $Z$, so the total effect of $X$ on $Z$ is $r_{XY} r_{YZ}$.  The second is the analysis $Z \sim X$.  In this analysis, the coefficient for $X$ will indicate the same causal impact $r_{XY} r_{YZ}$, and because there are no other pathways between $X$ and $Z$ except mediation through $Y$, this value is equivalent to $r_{XZ}$.

The overall causal impact of $X$ on $Z$ can be found by eliminating the $Y$ from the two previous regression equations, yielding: $Z = (\beta_{XY}\beta_{YZ})X$ = ($r_{XY}r_{YZ})X$.  The coefficient $r_{XY}r_{YZ}$ represents the causal impact of $X$ on $Z$.  Therefore, the analysis $Z \sim X$ will produce the correct causal coefficient.

In the idealized chain $X \rightarrow Y \rightarrow Z$, controlling for $Y$ will block the causal relationship between $X$ and $Z$.  Analyzing the relationship of $X$ on $Z$ while controlling for $Y$ can be accomplished through the analysis $Z \sim X + Y$, which would yield a coefficient for $Y$ of $r_{YZ}$ and a coefficient for $X$ of zero.  This can be intuitively understood through the fire alarm example: controlling for the mediator ``smoke''  blocks the relationship between fire and alarm.  We could do so by very efficiently removing smoke from a room with a fume hood.  In this case, there will be no smoke in the room, whether or not a fire is present, and the alarm will not sound.  We could also hold smoke constant by filling the room with smoke using a fog machine.  In this case, the alarm will sound whether or not fire is present.  By holding the presence or absence of smoke constant, the causal link between fire and the alarm sounding is broken. Therefore, controlling for the mediator $Y$ will screen off information about the actual, indirect causal relationship of $X$ on $Z$.

Note that, even in this relatively simple case, the correct causal interpretation depends critically on having the correct causal diagram.  If the causal chain were actually $X \leftarrow Y \leftarrow Z$, then the causal coefficient $r_{XY}r_{YZ}$ would actually represent how changing $Z$ would impact $X$, not how changing $X$ would change $Z$.  Though it is often theoretically clear which factor is the cause and which is the effect, there are systems where determining causes and effects is non-trivial.


\subsection{Fork}

A fork is represented as $X \leftarrow Y \rightarrow Z$ (Fig. \ref{ForkDAG}).  Here, $Y$ is a common cause of both $X$ and $Z$. Therefore, $X$ and $Z$ are correlated because changes in $Y$ will lead to changes in both $X$ and $Z$, but this correlation does not reflect a causal relationship between $X$ and $Z$. An everyday example of a causal fork is $Shoe\; Size \leftarrow Age\; of\; Child \rightarrow Reading\; Ability$ \cite{freedmanbook}. Children with larger shoes tend to read at a higher level because they are older, but the relationship is not one of cause and effect. Giving a child larger shoes will not cause their reading ability to increase, nor will improving a child's reading ability cause their shoe size to increase.

Here, the causal diagram indicates that the causal impact of $Y$ on $X$ is represented through the equation $X = \beta_{YX}Y = r_{YX}Y$ and the causal impact of $Y$ on $Z$ is represented through $Z = \beta_{YZ}Y = r_{YZ}Y$.  The analysis  $Z \sim X$ will produce a coefficient for $X$ of $r_{XY} r_{YZ}$, but this indicates a \emph{non-causal} association between $X$ and $Z$.  $r_{XY} r_{YZ}$ reflects how $X$ and $Z$ are correlated through $Y$ but directly changing $X$ (through a method that does not change $Y$) will produce no effect on $Z$.   

To determine the correct causal coefficients, one can control for $Y$.  For instance, analyzing subsets of same-aged children, the remaining variations in shoe size and reading ability should be uncorrelated, reflecting that there is no association after the common cause is controlled for.  For the fork $X \leftarrow Y \rightarrow Z$, controlling for $Y$ through the analysis $Z \sim X + Y$ will produce a coefficient for $Y$ of $r_{YZ}$ and a coefficient for $X$ of zero.  These coefficients reflect the causal impact of $Y$ and $X$, respectively, on $Z$.  In causal analysis, a common cause of two variables (here, $Y$ is a common cause of $X$ and $Z$) is called a \emph{confounder} since, if uncontrolled, it confounds our ability to estimate the causal relationship between those two variables by contributing a non-causal association.  In the causal diagram representation, non-causal pathways, such as the one from $X$ to $Z$ through a fork, $X \leftarrow Y \rightarrow Z$, are called backdoor paths, and controlling for confounders closes these backdoor paths.

Note that the interpretation of which regression analysis yields an accurate estimate of causal impacts depends on the proposed causal structure.  For both the chain and the fork discussed, $Z \sim X$ will yield $Z = (r_{XY} r_{YZ})X$, and $Z \sim X + Y$ will yield $Z = (0)X + (r_{YZ})Y$.  Which coefficient is the causal coefficient, describing how intervening directly on $X$ can change $Z$: $r_{XY} r_{YZ}$ or zero? For the chain, the correct causal coefficient is $r_{XY} r_{YZ}$.  The appropriate causal analysis does not control for the mediator $Y$ since this will mask the actual causal relationship between $X$ and $Z$.  For the fork, the correct causal coefficient is zero.  The appropriate causal analysis does control for the confounder $Y$ since this will block the backdoor path that contributes a non-causal association between $X$ and $Z$.  This highlights the critical importance of constructing the correct causal diagram when estimating the causal impacts of one variable on another.  

\subsection{Collider}

A collider is represented as $X \rightarrow Y \leftarrow Z$ (Fig. \ref{ColliderDAG}).  Here, $Y$ is a common effect of both $X$ and $Z$.  In the idealized case depicted, $X$ and $Z$ are uncorrelated ($r_{XZ} = 0$) because there are no direct or backdoor paths connecting them.  

Here, the causal diagram indicates that the causal impact of $X$ on $Y$ and $Z$ on $Y$ is represented through the equation $Y = (\beta_{XY})X + (\beta_{ZY})Z$.  Because $X$ and $Z$ are uncorrelated in this idealized diagram, $ \beta_{XY}= r_{XY}$ and $\beta_{ZY}= r_{ZY}$ (if $X$ and $Z$ were correlated, the $\beta$'s could be computed with Eq. \ref{eq:regcoeff}).  These $\beta$'s are the correct causal coefficients and indicate how changing $X$ and $Z$ will affect $Y$.  $X$ and $Z$ do not become correlated through a collider, so the analyses $Z \sim X$ and $X \sim Z$ would both yield coefficients equal to zero, which correctly indicates the lack of causal association between them.  

\begin{figure}[h]
\centering
\includegraphics[width=8.6cm]{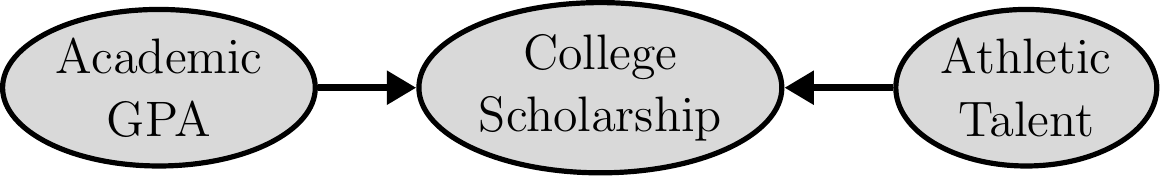}
\caption{Causal diagram illustrating the relationship between academic GPA, college scholarship, and athletic talent.}
\label{cb3}
\end{figure}

Here, controlling for $Y$ will produce a unique non-causal association: the  analysis $Z \sim X + Y$ will produce a non-causal coefficient for $X$ of $\frac{-r_{XY} r_{ZY}}{1-r_{XY}^2}$. That is, in the case that $X$ and $Z$ have positive causal impacts on $Y$, controlling for $Y$ will produce a negative non-causal association between $X$ and $Z$.  To see why this would be the case, consider an example of $Academic\; GPA \rightarrow College\; Scholarship \leftarrow Athletic\; Talent$ (Fig. \ref{cb3}). This causal diagram reflects that students can receive college scholarships based on either academic achievement or athletic talent (which, for the purposes of this example, we are imagining is largely uncorrelated with academic achievement). When we consider the subset of students who have been awarded a college scholarship (controlling for the collider), academic GPA will be anti-correlated with athletic talent.  For example, if a student receives a scholarship but they did not have a high GPA, it is more likely that they received a scholarship for playing sports.  Similarly, students without athletic talent are likely to have received a scholarship through a high academic GPA.  Here, controlling for the outcome opens a back-door path through the collider, revealing an association between causes that is present when considering a same-outcome subgroup but is not present when considering the entire population.

\section{Analyzing a simulated data set I: when you should and should not control for variables in causal inference}
\label{Controlforvariables}
Although one may be tempted to ``control for everything'' in quantitative analyses involving multiple measured variables, this approach does not necessarily produce the correct causal coefficients. For a chain, one should not control for mediators because this blocks the causal mechanisms that relates two variables. For a fork, one should control for confounders (common causes) to block non-causal associations between variables. One should not control for a collider (a common effect) because it conditions on an outcome that produces a non-causal association between the causes of that outcome. In sum, when seeking to produce an accurate estimate for the causal impact of $X$ on $Z$, one should control for confounders but not mediators or colliders. 

\begin{figure}[h]
\centering
\includegraphics[width=5.0cm]{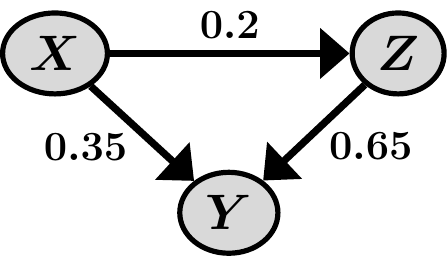}
\caption{Causal diagram depicting the relationships among variables $X$, $Y$, and $Z$, where $X$ causes $Z$, and $X$ and $Z$ jointly cause $Y$ with corresponding causal coefficients indicated on the arrows.}
\label{DAGbias}
\end{figure}

Though simply stated, the application of these rules can become more complex as the causal diagram becomes more complex.  To demonstrate these applications, we created a simulated data set based on the causal diagram shown in Fig. \ref{DAGbias}. The simulations were conducted using RStudio \cite{Rstudio}. First, a standardized, normal variable $X$ (mean = 0, standard deviation = 1) with $N = 10,000,000$ counts was simulated. Then, $N$ counts for a new variable $Z$ were computed from $X$ such that $Z$ would be a standardized variable where the regression analysis $Z \sim X$ would yield the best-fit line $Z= (0.20)*X$. Finally, $N$ counts for $Y$ were computed from $X$ and $Z$ such that $Y$ would be a standardized variable where the regression analysis $Y \sim X + Z$ would yield the line $Y = (0.35)*X+(0.65)*Z$. This step-wise simulation procedure followed the causal pathways in Fig. \ref{DAGbias}: $X$ determines $Z$, and then $X$ and $Z$ together determine $Y$.  Simulating the data in this way created a data set {$X$, $Y$, $Z$} where the correct causal diagram and the causal coefficients associated with each directed arrow are known (Fig. \ref{DAGbias}).  

Next, we demonstrate correct (and incorrect) analyses for determining the magnitude of causal effects between variables depending on the causal structure.  We show how the results determined from analyzing the simulated data match the causal structure depicted in Fig. \ref{DAGbias}. This will also illustrate how the path coefficients in Fig. \ref{DAGbias} can be used to determine the magnitudes of various causal and non-causal associations.

\subsubsection{Rule: do not control for mediators}
\label{MB}

Consider an analysis aiming to determine the causal impact of $X$ on $Y$. Fig. \ref{DAGbias} shows that this total causal effect is the sum of the direct path $X \rightarrow Y$ and the indirect path $X \rightarrow Z \rightarrow Y$.  This means that the total causal effect is $0.48$: the sum of the direct effect $0.35$ and the indirect effect $(0.2)(0.65)=0.13$.  This indicates that changing $X$ by +1 SD would produce a change in $Y$ of +0.48 SD.

Conceptually, the relationship between total, direct, and indirect causal effects can be understood as being akin to how total and partial derivatives are connected through the chain rule.  Consider the function $y = f(x,z(x))$, where $x, y, z \in \mathbb{R}$. This function is analogous to the causal diagram in Fig. \ref{DAGbias}, since $y$ depends on $x$ and $z$, while $z$ itself also depends on $x$.  The total derivative $\frac{dy}{dx}$ can be written as:

\begin{equation}
    \frac{dy}{dx} = \frac{\partial y}{\partial x} + \frac{\partial y}{\partial z} \frac{\partial z}{\partial x}
\end{equation}

Analogously, $\frac{dy}{dx}$ represents the total effect of $x$ on $y$, $\frac{\partial y}{\partial x}$ represents the direct effect of $y$ on $x$ when keeping $z$ constant, and $\frac{\partial y}{\partial z} \frac{\partial z}{\partial x}$ represents the indirect effect of $y$ on $x$ that is mediated through $z$.  
The correct analysis for determining the total causal effect is $Y \sim X$.  Because $Z$ is a mediator in the indirect causal path, it should not be controlled for, as doing so will block this causal path. The linear regression analysis of the simulated data results in the equation:

\begin{equation}
Y = (0.48)*X\label{2a}
\end{equation}

This yields the total causal effect of $X$ on $Y$. Controlling for $Z$ means performing the analysis $Y \sim X + Z$, which yields:

\begin{equation}
 Y = (0.35)*X + (0.65)*Z \label{2b}
\end{equation}

Controlling for the mediator blocks the indirect causal effect, leaving only the direct causal effect, 0.35, as the coefficient for $X$. 

\subsubsection{Rule: control for confounders}
\label{ConfoundingBias}

How would one determine the causal impact of $Z$ on $Y$?  Now, $X$ is a confounder (common cause) of $Z$ and $Y$, so it should be controlled in the analysis.  The total causal effect of $Z$ on $Y$ is only the direct effect, 0.65.  The confounder creates a non-causal association of (0.2)(0.35) = 0.07 through the backdoor path $Z \rightarrow X \rightarrow Y$.  

The correct analysis for determining the total causal effect of $Z$ on $Y$ is $Y \sim Z + X$.  When applied to the simulated data, this analysis yields:

\begin{equation}
Y = (0.65)*Z + (0.35)*X\label{5b}
\end{equation}

Here, controlling for $X$ in the analysis means that the coefficient for $Z$ will be the causal coefficient, representing the total causal effect 0.65. However, if one does not control for $X$ by performing the analysis $Y \sim Z$, this yields an incorrect causal coefficient:

\begin{equation}
Y = (0.72)*Z\label{4b}
\end{equation}

Note that for linear regression with one standardized independent variable $Z$ and one standardized dependent variable $Y$, the regression coefficient equals $r_{YZ} = 0.72$.  Because $X$ was not controlled for, the backdoor path added the non-causal association 0.07 to the causal effect 0.65 to produce the regression coefficient 0.72.  This example illustrates the problem with unmeasured confounders. Because the confounders must be controlled for in the analysis to produce the correct causal coefficients, the existence of unmeasured confounders makes accurate causal analysis impossible. This is why observational study design should seek to measure all confounders or proxies for these confounders so that they can be controlled for to block the non-causal associations from backdoor paths.

\subsubsection{Rule: do not control for colliders}
\label{ColliderBias}

For $X$ and $Z$, $Y$ is a collider.  Because colliders should not be controlled in causal analysis, the analysis for determining the causal impact of $X$ on $Z$ should not control for $Y$.  Controlling for variable $Y$ creates a bias by establishing a negative correlation between independent variables $X$ and $Z$. This bias is called collider bias. To find the causal coefficient of $X$ on $Z$ (0.2), one should not control for $Y$. Doing so will condition on a collider, opening a non-causal association between $X$ and $Z$ through $Y$. To demonstrate this, first, we perform the regression analysis without controlling for $Y$, $Z \sim X$, which yields:

\begin{equation}
    Z =(0.20)*X\label{2b}
\end{equation}

This analysis gives the correct causal coefficient for $X$, 0.2. On the other hand, controlling for the collider $Y$ through the analysis $Z \sim X + Y$, yields:
\begin{equation}
    Z =(-0.19)*X + (0.81)*Y \label{3b}
\end{equation}

The coefficient for $X$ becomes negative, reflecting the fact that controlling for $Y$ in this analysis has opened an additional negative, non-causal association between $X$ and $Z$. This is an extreme example showing how controlling for a collider can even flip the sign of a regression coefficient, and naive interpretation of these analyses could produce different conclusions about the causal impact of $X$ on $Z$.

\subsubsection{Omitted Variable Bias}
\label{OMV}

Omitted variable bias \cite{gelman2006data,clarke2005phantom,clarke2009return,riegg2008causal} is one term used to describe the change in regression coefficients when the analysis does not control for other variables \cite{walsh2021exploring}, an effect just demonstrated three times using simulated data. The general conditions for omitted variable bias are that (i) the omitted variable has a non-zero regression coefficient when predicting the dependent variable and (ii) the omitted variable is correlated with other independent variables used in the regression analysis.  

Although mathematically accurate, labeling this effect a ``bias" may suggest that no measured variables should be omitted in analyses where causal inference is the goal.  This is incorrect. Although mediators, confounders, and colliders all satisfy the two general conditions for omitted variable bias, only confounders should be controlled for in causal inference; mediators and colliders should be omitted in analysis. Controlling for mediators and colliders in analysis biases coefficients away from total causal effects.

\section{The importance of randomization in causal inference}

\begin{figure}
     \centering
     \begin{subfigure}[b]{0.45\textwidth}
         \centering
         \includegraphics[width=\textwidth]{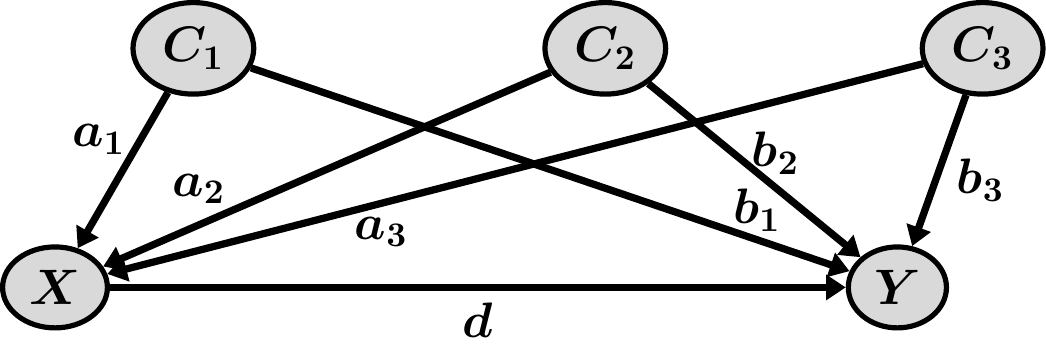}
         \caption{}
         \label{Randomization1}
     \end{subfigure}
     \hfill
     \begin{subfigure}[b]{0.45\textwidth}
         \centering
         \includegraphics[width=\textwidth]{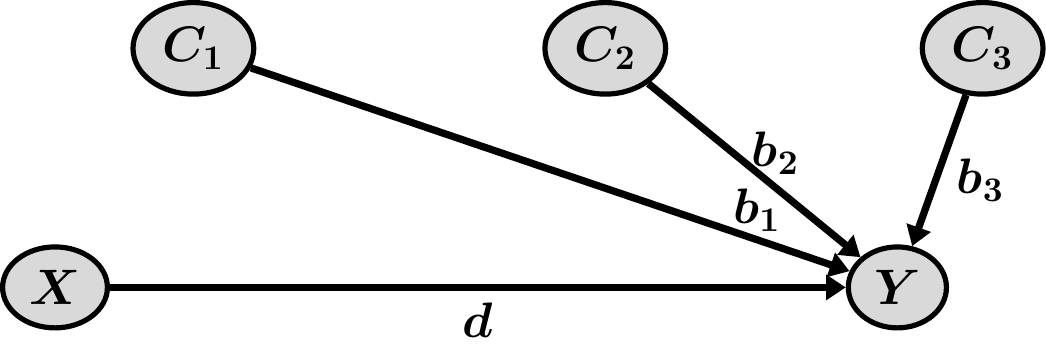}
         \caption{}
         \label{Randomization2}
     \end{subfigure}
        \caption{(a) Causal diagram illustrating the relationship between $X$ and $Y$, where $X$ causes $Y$. $C_1,C_2$, and $C_3$ are common causes of both $X$ and $Y$. The labels on the edges represent the causal coefficients. (b) A randomized experiment (or intervention) on $X$ breaks the causal dependence of confounders on $X$.}
        \label{fig:Randomization}
\end{figure}

One benefit of using DAGs to create causal diagrams is that the diagrams can concretely represent familiar issues in causal inference.  For example, causal diagrams can illustrate why randomized controlled trials (RCTs) \cite{pearl2018book,hutchison2010guide,torgerson2013every,hedges2018randomised}, where research participants are randomly assigned to a control or intervention group, are considered a ``gold standard'' for accurately determining the causal impact of one factor on another.  Consider the case in Fig. \ref{Randomization1}, where $X$ has a direct causal impact on $Y$, $X \rightarrow Y$, and multiple confounds $C_1$, $C_2$, and $C_3$ – common causes of $X$ and $Y$ – exist.  

If we observe these variables in situ, the regression analysis that will produce the correct causal coefficient of $X$ on $Y$, $d$, is $Y \sim X + C_1 + C_2 + C_3$.  The regression analysis $Y \sim X$ will produce a coefficient for $X$ that is the sum of the direct causal effect of $X \rightarrow Y$, $d$, and the non-causal associations due to the three backdoor paths $X \leftarrow C_1 \rightarrow Y$, $X \leftarrow C_2 \rightarrow Y$, and $X \leftarrow C_3 \rightarrow Y$.  Using the coefficients in Fig. \ref{Randomization1}, this coefficient will be $d + a_1 b_1 + a_2 b_2 + a_3 b_3$.  This is another example of why controlling for confounders in causal analysis matters.

Randomization is another way to deal with confounders without explicit measurement and control.  If experimenters can randomly assign participants to $X$ (such as if it is an instructional approach or learning activity that either be given or not given to students), this will break the causal dependence on confounders since the presence of $X$ will no longer depend on $C_1$, $C_2$, or $C_3$.  In this new diagram (Fig. \ref{Randomization2}), both the analysis $Y \sim X$ and $Y \sim X + C_1 + C_2 + C_3$ produce the same coefficient for $X$, $d$.  Therefore, randomization theoretically removes the need to control for, measure, or even be aware of confounders.

\section{Analyzing a simulated data set II: how multiple causal models can interpret the same data set}

In section \ref{Controlforvariables}, because we were privy to the exact causal process through which the data were simulated, we were certain of how the variables were causally related. However, this is rarely (if never) the case.  Although there may be cases where common-sense and theoretically-motivated causal models are not contentious, there are also instances where the exact causal model is uncertain or multiple causal models may be theoretically plausible. An important point is that quantitative statistics of model fit, though good at quantifying the predictive value of a statistical model, cannot be used to determine the correct causal model.  That is because the model fit is about explaining variance, but it does not specify whether that explanation indicates cause, effect, or non-causal association.

As a clear example of why model fit does not equal causal validity, consider the simulated data set represented by Fig. \ref{DAGbias}. As explained previously, in determining the causal impact of $X$ on $Y$, $Z$ should not be controlled because $Z$ is a mediator of this causal impact. Therefore, the correct causal analysis is $Y \sim X$.  However, if we make the choice of whether or not to control for $Z$ based on which regression model produces the highest $R^2$ fit, we will reach the wrong answer. The correct causal analysis $Y \sim X$ has an $R^2 = 0.23$, and the incorrect causal analysis $Y \sim X + Z$ has an $R^2 = 0.63$. The reason is that including the mediator $Z$ in the prediction of $Y$ explains additional variance compared to when only $X$ is used to predict $Y$, even though controlling for that mediator obscures the causal coefficient of $X$ on $Y$. Although including mediators and colliders in regression models can provide a greater predictive fit by explaining a greater proportion of variance in the dependent variable, it can also bias regression coefficients away from estimates of the total causal effect.

\begin{figure*}[htbp]
     \centering
     \begin{subfigure}[b]{0.25\textwidth}
         \centering
         \includegraphics[width=\textwidth]{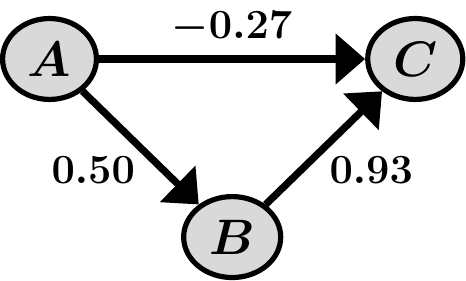}
         \caption{Model 1}
         \label{Model1DAG}
     \end{subfigure}
     \hfill
     \begin{subfigure}[b]{0.25\textwidth}
         \centering
         \includegraphics[width=\textwidth]{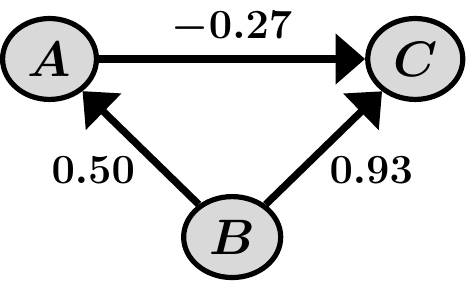}
         \caption{Model 2}
         \label{Model2DAG}
     \end{subfigure}
     \hfill
     \begin{subfigure}[b]{0.25\textwidth}
         \centering
         \includegraphics[width=\textwidth]{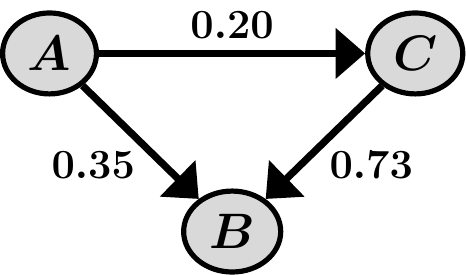}
         \caption{Model 3}
         \label{Model3DAG}
     \end{subfigure}
     \hfill
     \begin{subfigure}[b]{0.25\textwidth}
         \centering
         \includegraphics[width=\textwidth]{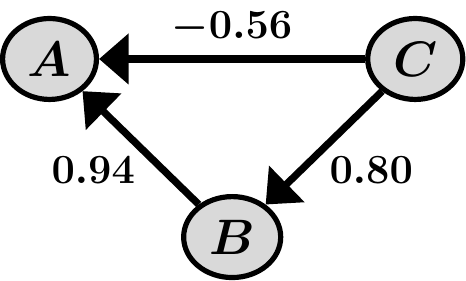}
         \caption{Model 4}
         \label{Model4DAG}
     \end{subfigure}
     \hfill
     \begin{subfigure}[b]{0.25\textwidth}
         \centering
         \includegraphics[width=\textwidth]{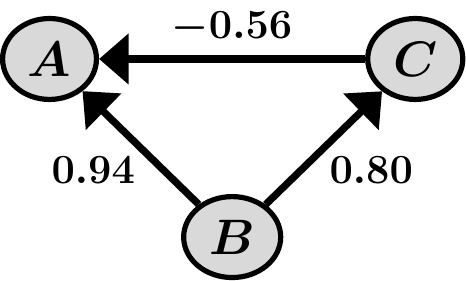}
         \caption{Model 5}
         \label{Model5DAG}
     \end{subfigure}
     \hfill
     \begin{subfigure}[b]{0.25\textwidth}
         \centering
         \includegraphics[width=\textwidth]{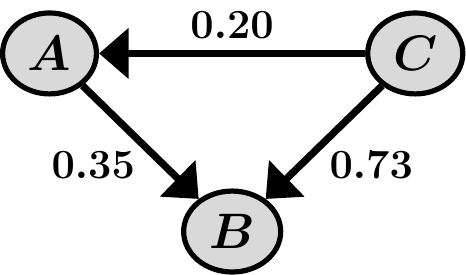}
         \caption{Model 6}
         \label{Model6DAG}
     \end{subfigure}
        \caption{Six acyclic causal models that fit the same set of correlations between the variables in Table \ref{covariance matrix}.}
        \label{Multipleequivalentcausalmodels}
\end{figure*}

Likewise, the existence of non-zero coefficients associated with an arrow in the causal diagram does not prove the validity of the model. To illustrate this, we simulated a data set of standardized variables ${A,B,C}$ that followed the correlations in table \ref{covariance matrix}. As with observational studies, the underlying causal model relating the different variables is not explicitly known (unlike the previously discussed simulated data set).  We analyzed the data using six different causal models (Fig. \ref{Multipleequivalentcausalmodels}).  These six models are all the ones allowed when considering models where all pairwise direct links exist and omitting the cyclic models which are disallowed.  For each model, the diagram represents the analyses required to find the path coefficients by considering which independent variables have direct arrows going into dependent variables.  For instance, in Model 1, $B$ only has a direct arrow pointing into it coming from $A$, so the path coefficient for $A \rightarrow B$ can be determined through the analysis $B \sim A$, which will yield the regression line $B = (0.5)A$.  $C$ has direct arrows pointing into it from both $A$ and $B$, so these path coefficients are determined by the analysis $C \sim A + B$, which will yield the regression line $C = (-0.27)A + (0.93)B$.  Because $A$ has no incoming arrows, it is not the dependent variable in any analysis.

Although all models find non-zero path coefficients, they make different predictions about how changing one variable will change another.  For example, consider the question, ``what is the causal effect of changing $A$ on $C$?''  Model 1 gives a direct effect of -0.27, an indirect effect of (0.50)(0.93) = 0.47, and a total causal effect of $(-0.27)+(0.47)=0.2$.  Model 2 gives a total effect of -0.27, which is solely attributed to a direct effect.  Model 3 gives a total effect of 0.20, which is solely attributed to a direct effect.  Models 4, 5, and 6 give an effect of zero since $C$ is the cause and $A$ is the effect, and changing $A$ directly will not change $C$.  

Although Models 1 and 3 give the same total causal effect, this degeneracy is broken when considering how holding $B$ fixed will change this causal effect.  In model 1, holding $B$ fixed will block the indirect effect, changing the total effect to -0.27.  In model 3, holding $B$ fixed will not affect the causal relationship between $A$ and $C$, so the total effect will remain 0.20.  Note that holding $B$ fixed is different than controlling for $B$ in analysis.  An intervention that holds $B$ fixed while changing $A$ would break the causal paths $A \rightarrow B$ and $C \rightarrow B$, while controlling for $B$ in the analysis would open a non-causal association between $A$ and $C$.  

Another issue is how the different models have different causal implications, even if the path coefficients are numerically identical.  For example, consider Models 1 and 2, which have the same numerical path coefficients and differ only in how the link between $A$ and $B$ is modeled: either $A \rightarrow B$ or $A \leftarrow B$.  Both models give a direct effect for $A \rightarrow C$ of $-0.27$.  In model 1, the path through $B$ is causal.  $B$ is a partial mediator through the path $A \rightarrow B \rightarrow C$, which represents an indirect, causal effect of $(0.5)(0.93) = 0.47$.  In model 2, the path through $B$ is non-causal.  $B$ is a confounder, so the path $A \leftarrow B \rightarrow C$ represents a non-causal association of magnitude $(0.5)(0.93) = 0.47$.  Therefore, although the choice of $A \rightarrow B$ or $A \leftarrow B$ has no impact on the direct path coefficients computed, it does have an impact on the causal implications of the model.

\begin{table}[htbp]
\centering
\begin{tabular}{cccc}
    \toprule
    & A & B & C \\
    \midrule
    A & $1.00$ & $0.50$ & $0.20$ \\
    B & $0.50$ & $1.00$ & $0.80$ \\
    C & $0.20$ & $0.80$ & $1.00$ \\
    \bottomrule
\end{tabular}
\caption{Correlation matrix for variables A, B, and C.}
\label{covariance matrix}
\end{table}

This simulated example shows the importance of justifying the \textit{a priori} causal model and its underlying assumptions about cause and effect.  Researchers have the freedom to choose any causal model and apply it to the data, and the choice of model changes the conclusions that will be reached.  The choice of model can even change the sign of a causal effect, as demonstrated in the simulated example, flipping its conceptual contribution to an effect.  Just like an ansatz, the causal model is a guess – however theoretically or empirically justified – about the causal relationships among a system of variables.  However, finding a model that fits the data is not proof that the ansatz was correct in this case.  In fact, neither statistical goodness-of-fit nor non-zero path coefficients offer evidence supporting the causal validity of one model over another. The results are only as valid as the researchers' original causal assumptions embodied in the proposed causal diagram.  As the number of relevant and collected variables grows, the number of possible causal models also grows, increasing the possibility that researchers have chosen the wrong model and reached the wrong causal conclusions.  

Since the results of causal analysis under a proposed model cannot support the likelihood of that proposed model over others, how can observational research proceed in fields like PER?  One way forward is to bridge observational and intervention studies, just as bridging theory and experiment has advanced knowledge in physics. Like theoretical models in physics, fitting observational data with theoretical causal models can make clear predictions of the outcomes of future interventions, motivating future intervention studies.  Like experiments in physics, intervention studies that directly manipulate causes and measure changes to effects can provide empirical data about which associations are causal and which are not that help support or falsify proposed theoretical models.  

Next, we will apply these causal inference methods to interpret prior work in PER. In doing so, we will provide an example of how observational studies can propose quantitative theoretical models that can be investigated through future intervention studies.

\section{Applying Causal Inference Principles to Prior PER studies}

Although PER often uses quantitative analysis to draw conclusions about causal impacts, the causal diagrams, assumptions, and analytic techniques discussed in this paper are rarely explicitly employed to justify and structure the analysis. Here, we apply these causal inference methods to make sense of prior work in PER, demonstrating how these methods can provide a unified language for understanding various issues in quantitative PER.  

\subsection{Example: Omitted Variable Bias in PER}

\begin{figure*}[htbp]
\centering
\includegraphics[width=12.0cm]{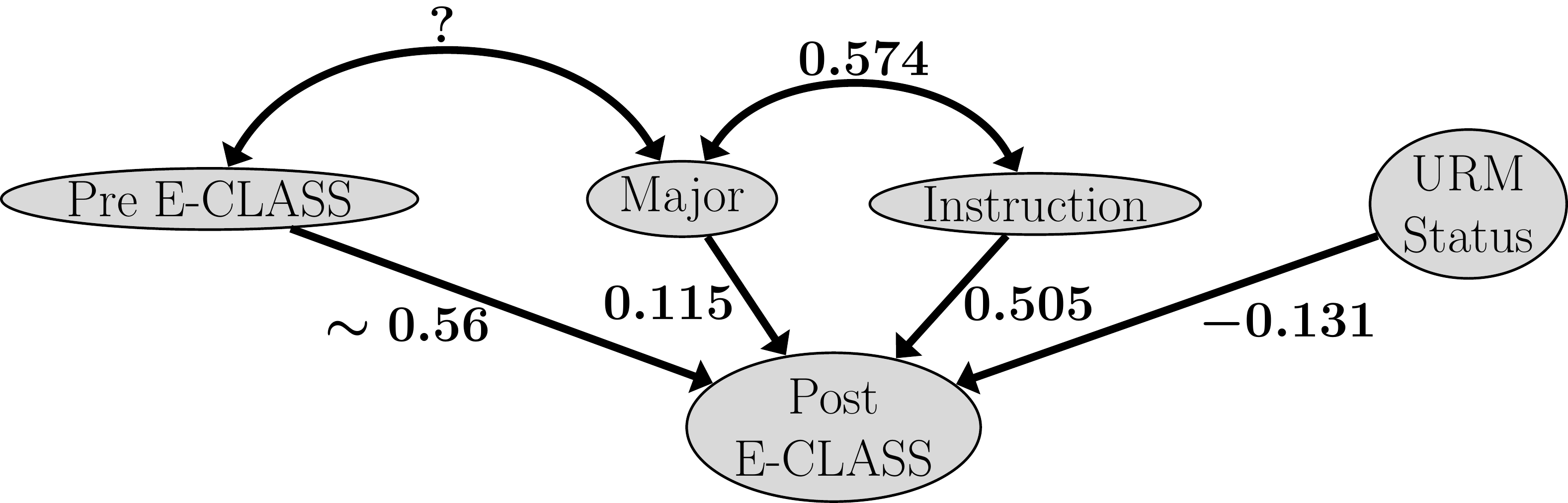}
\caption{A proposed causal diagram representing the causal structure of how pre E-CLASS scores, major, instruction, and URM status predicts the post E-CLASS scores. }
\label{fig:CausaldiagramsWalsh}
\end{figure*}

Walsh et al. \cite{walsh2021exploring} explored the effects of omitted variable bias through data from a quasi-experimental study. The study investigated students' attitudes towards experimental physics using pre and post E-CLASS survey measurements.  Sampled physics students experienced either ``transformed'' or ``highly traditional'' physics laboratory instruction and were coded as either intending to major in physics or intending to major in another science/engineering field.  Additionally, students' underrepresented minority (URM) status was collected.  The focus of their analysis was the magnitude of the omitted variable bias from omitting instruction type from the analysis. They create three regression models using different combinations of pre E-CLASS score, major (physics $=1$), instruction (transformed $= 1$), and URM status (URM $= 1$) to predict post E-CLASS score.  Using the correlation and regression results given in the paper, we propose a causal diagram for these variables (Fig. \ref{fig:CausaldiagramsWalsh}).  Because the addition of URM status and instruction has a very small effect on the regression coefficients for the pretest and major, we approximate the direct effects between these two groups as zero.  

The key question is, ``what is the causal relationship between major and instruction type?''  Without more knowledge of how instruction type is assigned to students, it is impossible to know what the most plausible causal relationship is.  In our diagram, we represent this connection as a non-causal association: $Major \leftrightarrow Instruction$.  It is conceptually equivalent to the notation: $Major \leftarrow U \rightarrow Instruction$, where \emph{U} is the unmeasured common cause of major and instruction. For instance, different types of instruction may be randomly assigned to different lab sections, and students may be blind to which sections are associated with each type of instruction.  In this case, the association would be purely non-causal since there would be no causal mechanism for students' major to influence which lab instruction they receive.  A second plausible causal relationship is that students' major may influence the type of lab instruction they receive: $Major \rightarrow Instruction$. For instance, the transformed lab instruction may be officially associated with lab sections for majors such that students are officially advised to enroll in different lab sections by major.  The transformed lab may also be messaged as ``more advanced'' or ``for physics majors'' in other ways that preferentially attract physics majors. The causal interpretation of this alternative model will be explored later.

Table I in \cite{walsh2021exploring} describes the results from the different regression models used. Model 1 performs the analysis $Post\; E-CLASS \sim\; Major+ Pre \;E-CLASS$. Using our diagram, we can see that controlling for $Pre\; E-CLASS$ blocks the backdoor path $Major \leftrightarrow Pre\; E-CLASS \rightarrow Post\;E-CLASS$, but because this analysis does not control for Instruction, the backdoor path from $Major \leftrightarrow Instruction \rightarrow Post\; E-CLASS$ is open. Using this diagram, we can determine that the regression coefficient for major will not be the correct causal coefficient. This regression coefficient, 0.405, will be the sum of the causal direct effect, 0.115, and the non-causal backdoor association $Major \leftrightarrow Instruction \rightarrow Post\; E-CLASS$, (0.574)(0.505) = 0.290.  This is approximately equal to the regression coefficient for Major computed in \cite{walsh2021exploring}, given for Model 1 in Table I.

Model 2 is the regression analysis $Post\; E-CLASS \sim Major + Instruction + Pre\; E-CLASS$.  This analysis controls for Instruction, blocking the previously open non-causal backdoor path $Major \leftrightarrow Instruction \rightarrow Post\; E-CLASS$.  Now, the regression coefficients in this analysis will match the direct, causal effects in Fig. \ref{fig:CausaldiagramsWalsh}: 0.115 for major, 0.505 for instruction, and 0.56 for pre E-CLASS. These values match those given in Table I of \cite{walsh2021exploring} (though not exactly for pre E-CLASS since we made the approximation that there is no correlation between pre E-CLASS and instruction that is unexplained by major). This illustrates how causal diagrams can provide one model that explains the results of multiple possible regression analyses while also encoding the causal assumptions of the researchers.  Model 3 in Table I of \cite{walsh2021exploring} describes the regression analysis $Post\; E-CLASS \sim Major + URM\; status + Pre\; E-CLASS$.  With the backdoor path between Major and Instruction open again, the coefficient for Major will become similar to that in model 1.  Because the pre E-CLASS and major coefficients remain similar to the model 1 values, model 3 shows that URM status has a very small correlation with pre E-CLASS or major, which we approximate as zero.

What causal inferences can we make from the causal diagram in Fig. \ref{fig:CausaldiagramsWalsh}?  Under the theoretical assumption that there is no causal association between major and instruction ($Major \leftrightarrow Instruction$), the type of instruction that students receive should not affect or be affected by their intended major. Therefore, one causal inference represented by this causal model is that experiencing the transformed lab instruction would increase students' average post E-CLASS by an amount corresponding to a standardized coefficient of 0.505 over the traditional lab instruction. Here, major is a confounder, so it must be controlled for to close a non-causal pathway between instruction and post-E-Class. If major was not controlled for, such as through the regression analysis $Post\; E-CLASS \sim Instruction + Pre\; E-CLASS$, the instruction coefficient would be $0.571 = 0.505 + (0.574)(0.115)$, overestimating the causal coefficient by $(0.574)(0.115) = 0.066$ through the non-causal backdoor path $Instruction \leftrightarrow Major \rightarrow Post\; E-CLASS$.

What causal inference can be drawn about major?  Intended major is a proxy measure for factors that attract students to physics over engineering and other sciences, including academic preparation, interest, etc.  These factors are hidden in the diagram as the common causes of pre E-CLASS and major, represented by $Pre\; E-CLASS \leftrightarrow Major$. For students with the same pre E-CLASS score and experiencing the same lab instruction intending to major in physics will increase students' average post E-CLASS score by an amount corresponding to a standardized coefficient of 0.115 over those intending to major in other science/engineering fields. A causally ridiculous conclusion would be that universities should change all physics students' intended majors to physics in the university registration system because this would improve their experimental physics attitudes after instruction. To explain why this is a formally incorrect conclusion from the causal diagram, doing so would not make major a good proxy for the relevant underlying student factors, diminishing its association with the other variables.  In the extreme case where the university makes all students become physics majors, major would have zero association with any other variable. 

In contrast to this interpretation of major, we now consider the alternative causal model where instruction partially mediates the causal effect of major: $Major \rightarrow Instruction \rightarrow Post\; E-CLASS$.  If being a major increases the chances that one is enrolled in the transformed lab instruction course, then the transformed lab instruction can be considered part of the ways that major effects post E-CLASS score.  Now, the total causal effect of intending to major in physics would be 0.40, which includes the direct effect of major of 0.115, associated with the unmeasured student factors related to major choice (like academic preparation, interest, etc.) and the indirect effect of physics majors being preferentially guided into the transformed lab instruction and this lab instruction impacting students' experimental physics attitudes. 

Although Walsh et al. \cite{walsh2021exploring} do not explicitly propose a causal interpretation of the 0.560 correlation between major and instruction, the causal diagram and associated rules for causal inference make it clear why this specification is important. While a general focus on omitted variable bias highlights how including or omitting variables from the analysis can affect regression coefficients, these causal techniques highlight additional issues around how those coefficients should be interpreted for accurate causal inference.

\subsection{Example: Collider Stratification Bias through sampling in PER}

The issue of non-causal coefficients arising from controlling for colliders – commonly called collider stratification bias – has been explicitly discussed in many contexts \cite{greenland2003quantifying,whitcomb2009quantification,elwert2014endogenous,banack2015bad,sperrin2016collider,coscia2022avoiding,del2015collider,leite2020collider,sanni2013time,tonnies2022collider,holmberg2022collider,griffith2020collider,hernan2004structural,vanderweele2007directed,cole2010illustrating,elwert2014endogenous}. Weissman \cite{weissman2020gre} explicitly discusses this issue in the context of education research, explaining how collider stratification bias can arise when controlling for educational outcomes in analysis. We elaborate on another way that collider stratification bias can arise: through sampling.

\begin{figure}[htbp]
\centering
\includegraphics[width=8cm]{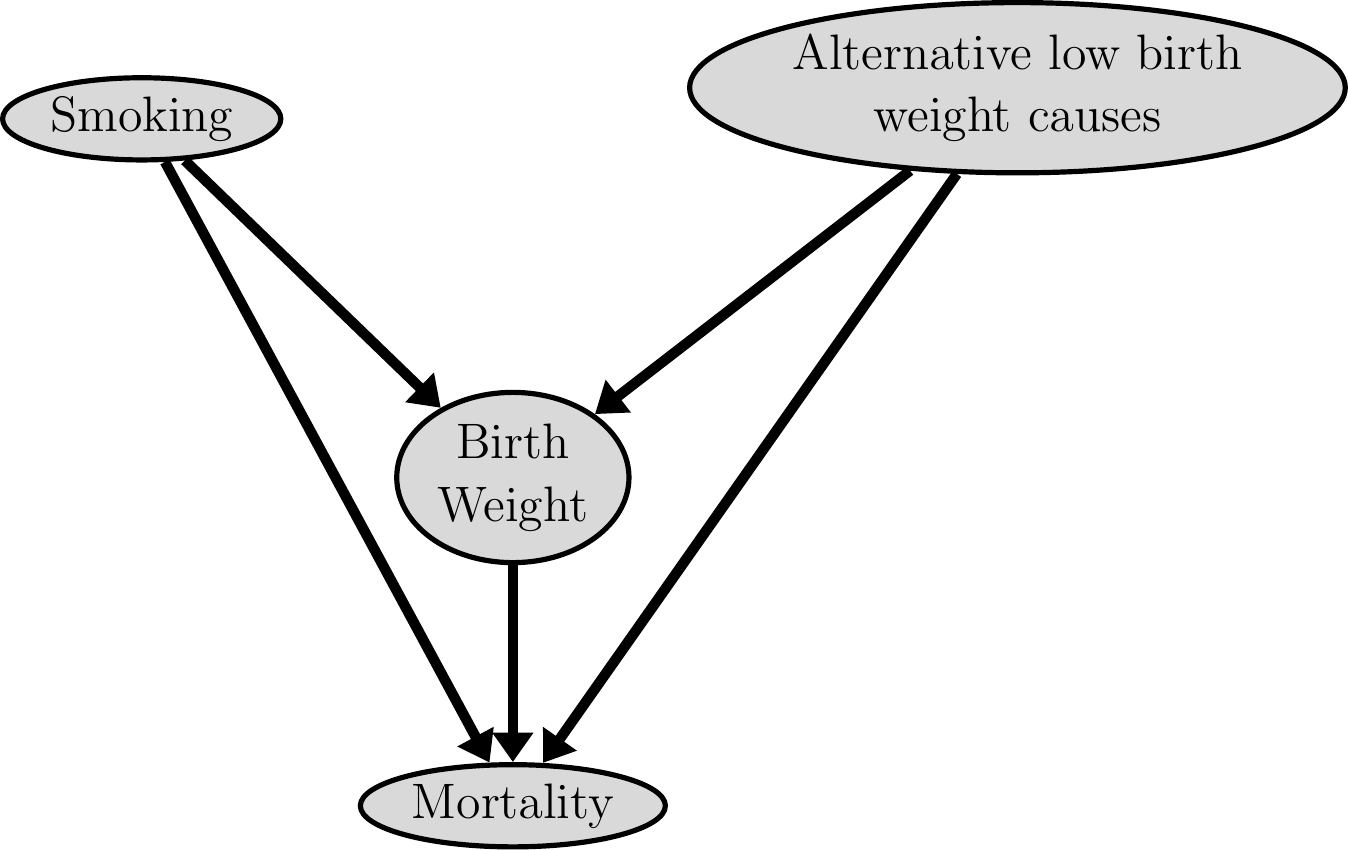}
\caption{Causal diagram illustrating the relationships between smoking, birth weight, alternative low birth weight causes, and mortality as seen in the study \cite{Yerushalmy1971}.}
 \label{DAGColliderStratbias}
\end{figure}

A study in the 1960s investigating the mortality of babies born with a low birth weight counterintuitively found that babies whose mothers were smokers had \emph{better} survival rates than babies of non-smoking mothers \cite{Yerushalmy1971}. Collider stratification bias was eventually used to explain why mothers should not be recommended to take up smoking while pregnant.  In this example, birth weight is a collider with multiple alternative causes. Smoking is one, but others also exist (such as birth defects).  Since the study only investigated low birth weight babies, the sampling conditioned on the collider, birth weight. The result is that smoking and alternative low birth weight causes have a non-causal association in the collected data set since low birthweight babies are likely to experience at least one of the causes, a smoking mother or an alternative low birthweight cause.  Babies who do not have a smoking mother are more likely to have alternative low birth weight causes, which may have even greater mortality rates than smoking.  Fig. \ref{DAGColliderStratbias} shows a causal model of these variables.  For the analysis, $Mortality \sim Smoking + Birth\;Weight$, the coefficient on smoking will be the sum of the effects of the direct path $Smoking \rightarrow Mortality$ and the non-causal backdoor path $Smoking \rightarrow Birth\; Weight \leftarrow Alternative\;Low\;Birth\;Weight\;Causes \rightarrow Mortality$.  If the non-causal backdoor path has a negative contribution greater in magnitude than the direct path, then the overall regression coefficient will be negative.  This explains how a smoking mother could predict a lower mortality rate than a non-smoking mother: because non-smoking becomes associated with other alternative causes of low birth weight with a higher mortality rate.

There are two potential ways one could address this collider stratification leading to non-causal coefficients of smoking on mortality. One would be to expand sampling to capture a representative distribution of birth weights. This would change the data set to not condition on the collider, closing this non-causal backdoor path through the collider. In this case, the analysis $Mortality \sim Smoking$ will produce a regression coefficient that represents the total causal effect (the direct effect of $Smoking \rightarrow Mortality$ plus the indirect effect of $Smoking \rightarrow Birth\; Weight \rightarrow Mortality$.  Another way to address the collider stratification would be to measure and control for alternative causes of low birth weight, like birth defects.  Although restricted sampling would still open the backdoor path through the collider (birth weight), controlling for birth defects and other alternative causes (which are common causes of birth weight and mortality) will close the non-causal confounding paths $Birth\;Weight \leftarrow Alternative\;Low\;Birth\;Weight\;Causes \rightarrow Mortality$.  In this case, the analysis $Mortality \sim Smoking + Alternative\;Low\;Birth\;Weight\;Causes$ (and conditioning on birth weight through restricted sampling) will produce coefficients that estimate the direct causal effects of $Smoking \rightarrow Mortality$ and $Alternative\;Low\;Birth\;Weight\;Causes \rightarrow Mortality$.  Although this removes contributions of non-causal backdoor associations from the regression coefficients, it also does not estimate the total causal impacts on mortality.  This is because birth weight is a partial mediator of the effects of smoking and alternative causes of low birth weight on mortality, and controlling for birth weight closes these mediation pathways.  

\begin{figure}[htbp]
\centering
\includegraphics[width=8cm]{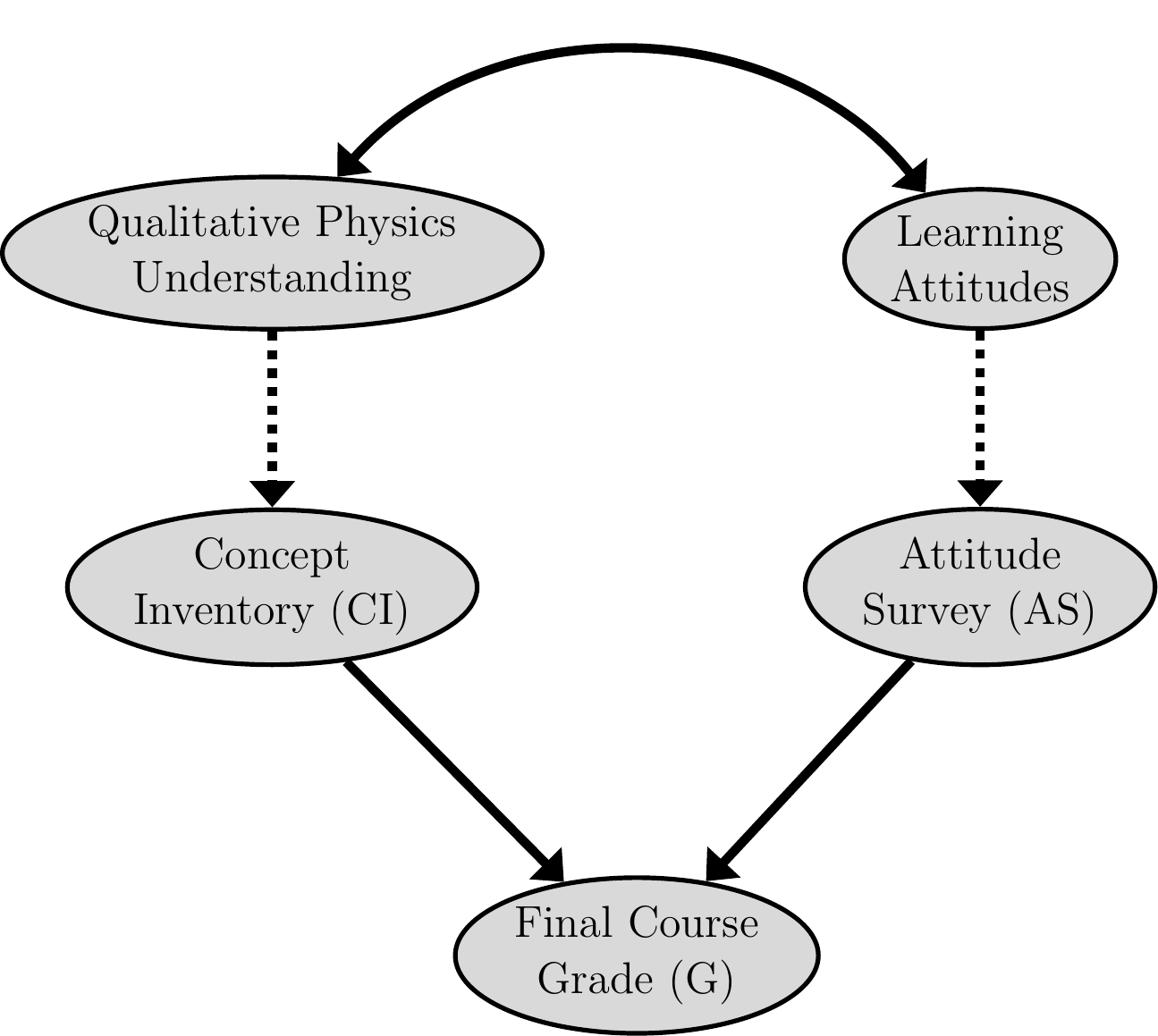}
\caption{A proposed causal diagram illustrating the causal structure of how concept inventory (CI) and attitude survey (AS) predict final course grade (G).}
\label{Colliderstartificationbias2}
\end{figure}

In PER, one data collection procedure where sampling creates collider bias is the completion of low-stakes, research-based surveys. Completion of these surveys, such as concept inventories or attitude surveys, during a physics course, is associated with final course grade: specifically, students with higher grades are more likely to complete these surveys \cite{nissen2018participation}.  For this reason, complete-case analysis, which removes participants with missing data from the analysis, will partially control for the final course grade. Consider the proposed causal model where a concept inventory (CI) and an attitude survey (AS) each serve as proxies for the qualitative physics understanding and learning attitudes that improve physics learning and performance as measured by final course grade (G) (Fig. \ref{Colliderstartificationbias2}). The partial control for the final grade partially opens the non-causal backdoor path through the collider $CI \rightarrow G \leftarrow AS$. Since we expect all causal coefficients to be positive, this backdoor path adds a non-causal negative contribution to the correlation between CI and AS. The expected impact is that measured correlations between CI and AS that do not address this collider stratification bias underestimate the strength of this correlation.  Biases associated with missing data have led to increased attention on data imputation techniques, like multiple imputation, for estimating the contributions of missing data in PER \cite{nissen2019missing}. Yet, just as with these causal inference methods, the accuracy of these methods depends critically on often unverifiable assumptions, in this case, about the nature of the missingness of the data and whether observed variables can adequately model the missing data.  

\subsection{Dealing with the cyclic nature of motivation and beliefs with linear models}

Because the causal models are created \emph{a priori}, the results of the analysis with these models are only as good as the assumptions that went into them.  Therefore, clarity about which variables are causes and which are effects determines the believability of the results. One area where the causal directions are manifestly bi-directional is between academic performance and motivation/beliefs.  

\begin{figure}[htbp]
\centering
\includegraphics[width=8cm]{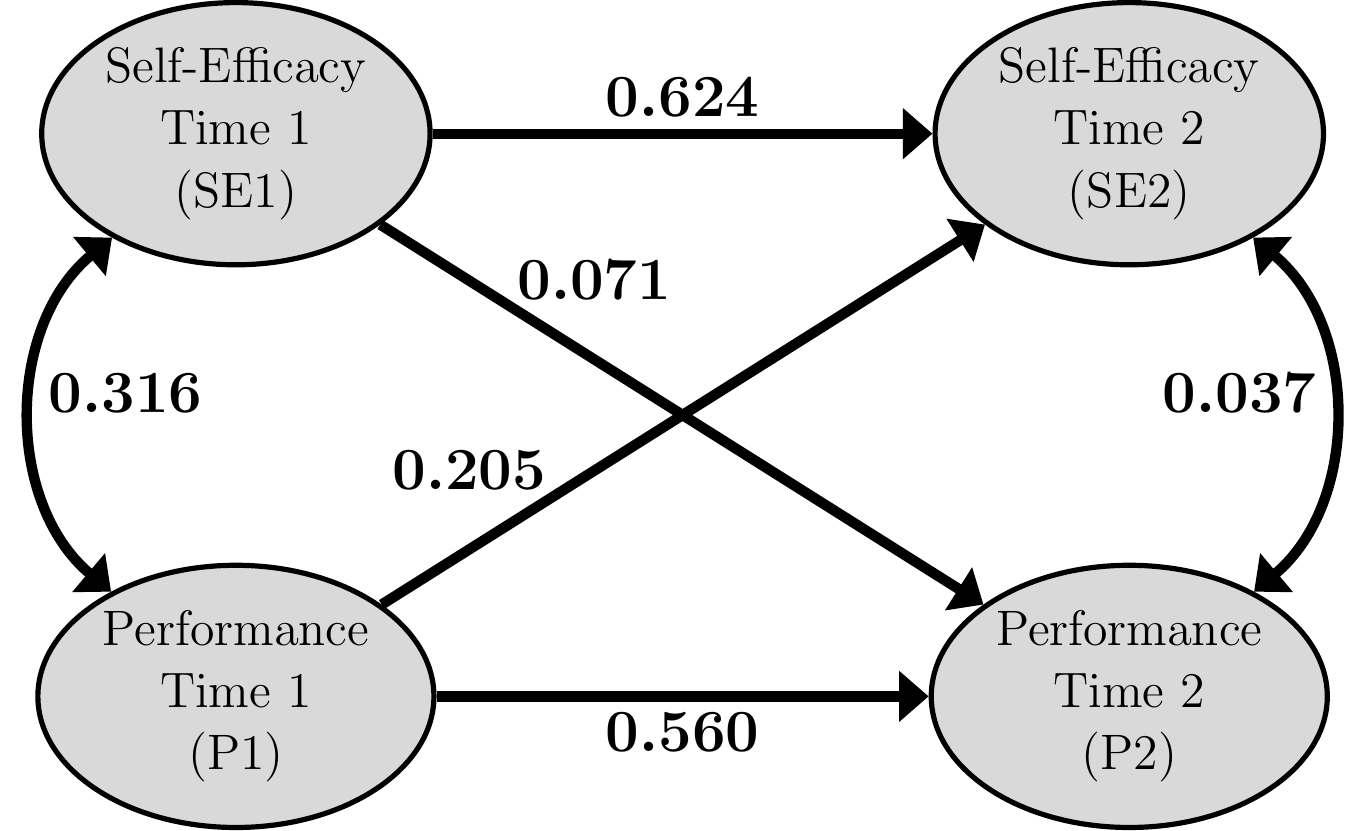}
\caption{This is a reproduction of Fig. 2 of \cite{talsma2018believe}. Causal diagram of the cross-lagged path model between self-efficacy and academic performance at time 1 and time 2.}
\label{CrossLaggedModel}
\end{figure}

To illustrate this, consider research on self-efficacy and academic performance. Although self-efficacy and academic performance are correlated with each other, which is the cause and which is the effect?  Although many researchers focus on one causal pathway over another ($SE \rightarrow \varname{performance}$ or $\varname{performance} \rightarrow SE$), from its conception, self-efficacy has been theorized to affect and be affected by behavior and performance \cite{bandura1986social,locke1997self,liem2008role,vogt2008faculty,jones2010analysis,honicke2016influence,vancouver2001changing,hattie2013international,pajares1997current,zimmerman2000self}.  Self-efficacy influences behaviors, such as whether or not people engage and persist in challenging tasks, which creates opportunities to increase learning and performance.  Reciprocally, experiencing mastery and success in performance is a strong predictor of future self-efficacy \cite{britner2006sources,usher2008sources,lent1991mathematics,klassen2004cross,matsui1990mechanisms}.  A sensible causal model between self-efficacy and academic performance would be cyclic \cite{multon1991relation, phan2012informational} (Fig. \ref{CrossLaggedModel}), representing the reciprocal relationship between the two factors. However, these cyclic causal diagrams are disallowed by formalism because the graphs must be acyclic.  In our own work on the relations between self-efficacy and performance \cite{boden2018role}, we have grappled with how to causally understand the quantitative relations between self-efficacy and physics performance in the absence of causal diagram methods. 

One way to conceptualize such reciprocal relationships is through longitudinal measurement and cross-lagged panel analysis.  As an example, Talsma et al. (2017) \cite{talsma2018believe} conducted a meta-analysis of longitudinal self-efficacy studies with a cross-lagged model where self-efficacy and performance correlated with each other at time 1, are both allowed to affect self-efficacy and performance, which remain correlated with each other, at time 2 (Fig. \ref{CrossLaggedModel}).  The longitudinal repeated measurements of performance and self-efficacy open up alternatives to cyclic diagrams.  The cross-lagged panel model also disentangles the effects of prior self-efficacy and prior performance, which are themselves correlated.  This causal diagram also clarifies the risks of simply associating self-efficacy at time 1 with academic performance at time 2. Talsma et al. \cite{talsma2018believe} report that this correlation is $r_{\text{SE1 - P2}} = 0.248$.  However, the analysis associated with this diagram shows that the causal effect $SE1 \rightarrow P2$ is only $0.071$ and that the rest of this correlation reflects a non-causal backdoor association $SE1 \leftrightarrow P1 \rightarrow P2$ of $(0.316)(0.560)=0.177$.  That is, the majority of this correlation reflects the fact that self-efficacy and performance at time 1 are correlated with each other and that the direct effect of $P1 \rightarrow P2$ is relatively large.  Neglecting this backdoor association in analysis overestimates the causal impact of self-efficacy on performance.  The cross-lagged diagram also clarifies that the correlation between self-efficacy and performance is mostly explained by the mechanism of their co-evolution over time.  Although $r_{\text{SE2 - P2}} = 0.312$, the direct non-causal association $SE2 \leftrightarrow P2$ only has a coefficient of $0.037$.  This indicates that an association of $0.275$ is explained through the backdoor paths, including SE1 and P1.  That is, most of the correlation between self-efficacy and performance is due to the fact that they both co-develop out of prior self-efficacy and performance.  

This example of self-efficacy shows how these causal methods can potentially clarify the muddy, reciprocal relations commonly theorized when considering relationships between academic performance and behavior with motivation, self-concept, and attitudes.  Although cross-lagged panel analysis illustrates the conceptual issues regarding reciprocal influences between variables, new methods have since been suggested that capture the same conceptual issues while relaxing some of the underlying assumptions required to produce accurate causal estimates \cite{mund2019beyond,hamaker2015critique,usami2019modeling}.

\subsection{Proposing an explicit role for causal modeling of observational data in PER: motivating future intervention studies}

The validity of the causal inferences drawn from quantitative analysis depends critically on the validity of the underlying causal model guiding analysis and interpretation.  This causal model, which can be represented explicitly with a DAG, is based on researchers' (explicit and/or implicit) theoretical understanding of the causal system.  The critical issue is that a researcher's underlying causal model cannot be ``verified'' by the quantitative results of fitting observational data to that model.  Intuitively, it may be appealing to interpret non-zero regression coefficients or extremizing quantitative metrics of model fit as evidence for a proposed causal model, and these quantitative results indicate the predictive power of the model, not the causal validity. Even a \emph{non-causally} correlated set of variables can produce non-zero regression coefficients and provide a good fit for predicting outcomes.  

In this journal, Weissman has called for explicit consideration of multiple plausible causal models for observational studies drawing causal inferences \cite{weissman2021policy}, which is especially relevant in cases with a large number of variables and possible causal connections between them and in cases with plausibly reciprocally developing student factors.  We agree with Weissman that this is a sensible call for considering alternative explanations in research.  As demonstrated previously, changing one's assumptions about whether a variable is a mediator, confounder, or collider can change quantitative causal estimates, as well as how the quantitative analysis should be conducted and interpreted. At the same time, we note the continued risk that researchers may incorrectly interpret explicit consideration and rejection of these alternative models as ``verifying'' their proposed causal model.  However sensible or convincing the theoretical arguments favoring one model over alternatives are, these theoretical arguments do not constitute empirical support for a model's causal validity.

Therefore, in addition to the consideration of alternative causal models, we propose an explicit goal for observational studies drawing causal inferences: proposing future intervention studies and predicting their outcomes. Just as physics theories motivate future experiments, any proposed causal model embodies a set of theoretical assumptions of how variables are causally related, and the causal estimates produced by applying those theoretical models to observational data are predictions of the effects of future interventions.  Framing causal inferences from observational studies as theory clarifies that these inferences are one set of proposed theoretical explanations for observed correlations.  Articulating these proposed causal models can motivate intervention studies that will investigate the impacts of manipulating proposed causes on effects.  Moreover, quantitative modeling provides quantitative predictions for how an intervention will affect other variables and relations between variables. A secondary benefit of this explicit framing of ``causal inference from observational data as theory'' is that it highlights and promotes the value of intervention studies. When interventions act on causes, they can break associations with confounders, eliminating non-causal backdoor paths and providing strong tests of proposed causal models.

\begin{figure*}[htbp]
     \centering
     \begin{subfigure}[b]{0.45\textwidth}
         \centering
         \includegraphics[width=\textwidth]{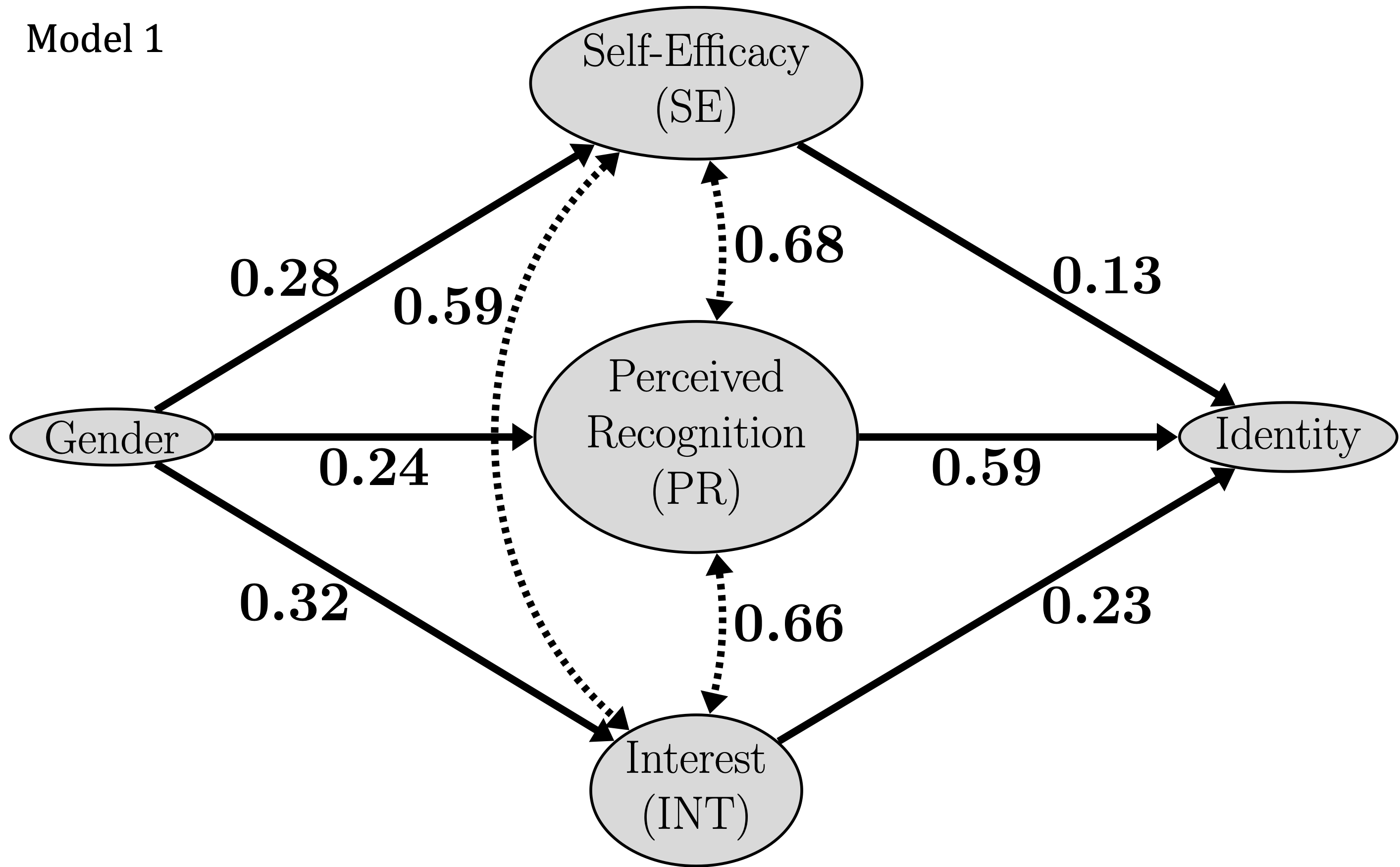}
         \caption{}
         \label{FIModel1}
     \end{subfigure}
     \hfill
     \begin{subfigure}[b]{0.45\textwidth}
         \centering
         \includegraphics[width=\textwidth]{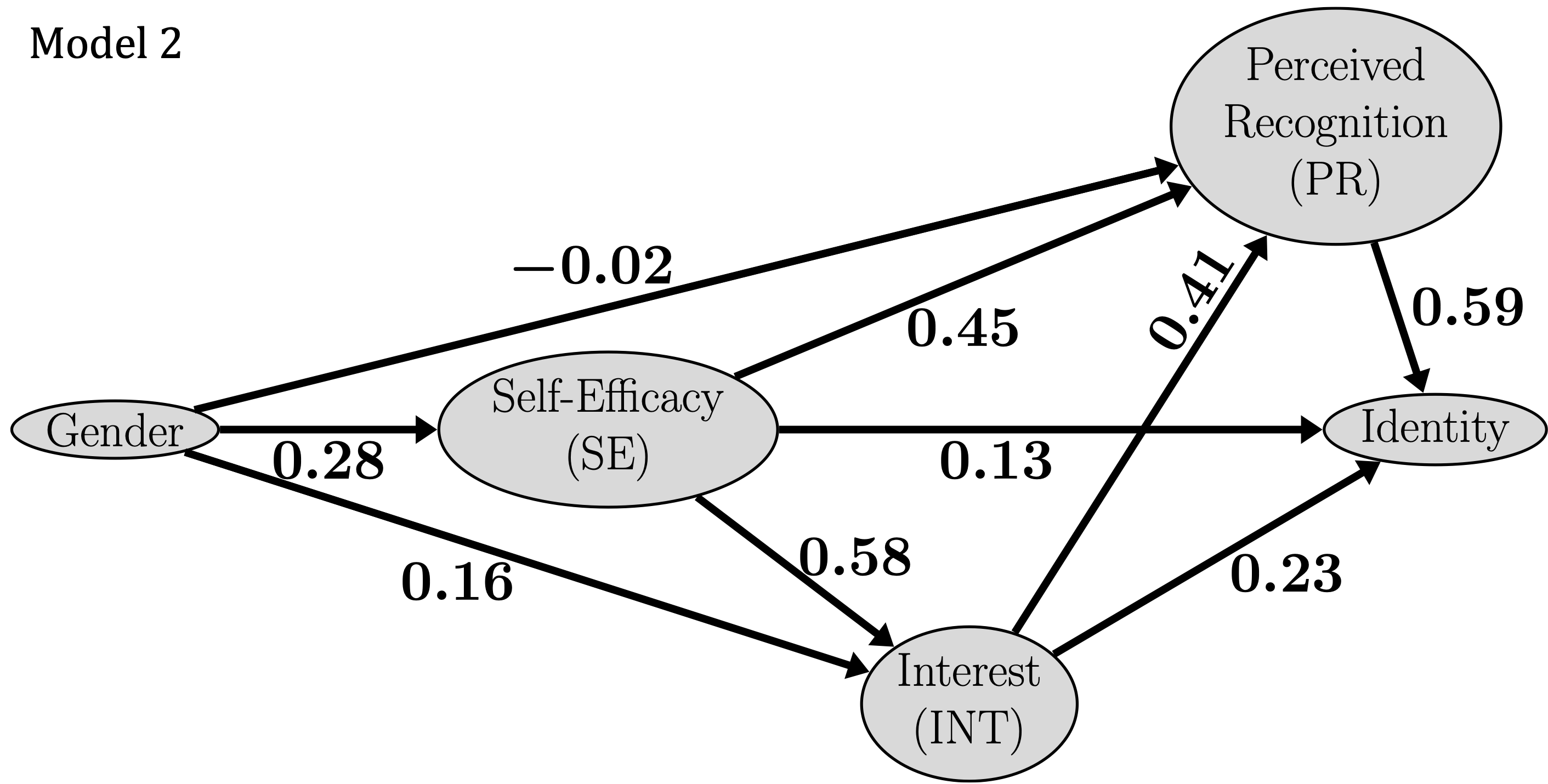}
         \caption{}
         \label{FIModel2}
     \end{subfigure}
     \hfill
     \begin{subfigure}[b]{0.45\textwidth}
         \centering
         \includegraphics[width=\textwidth]{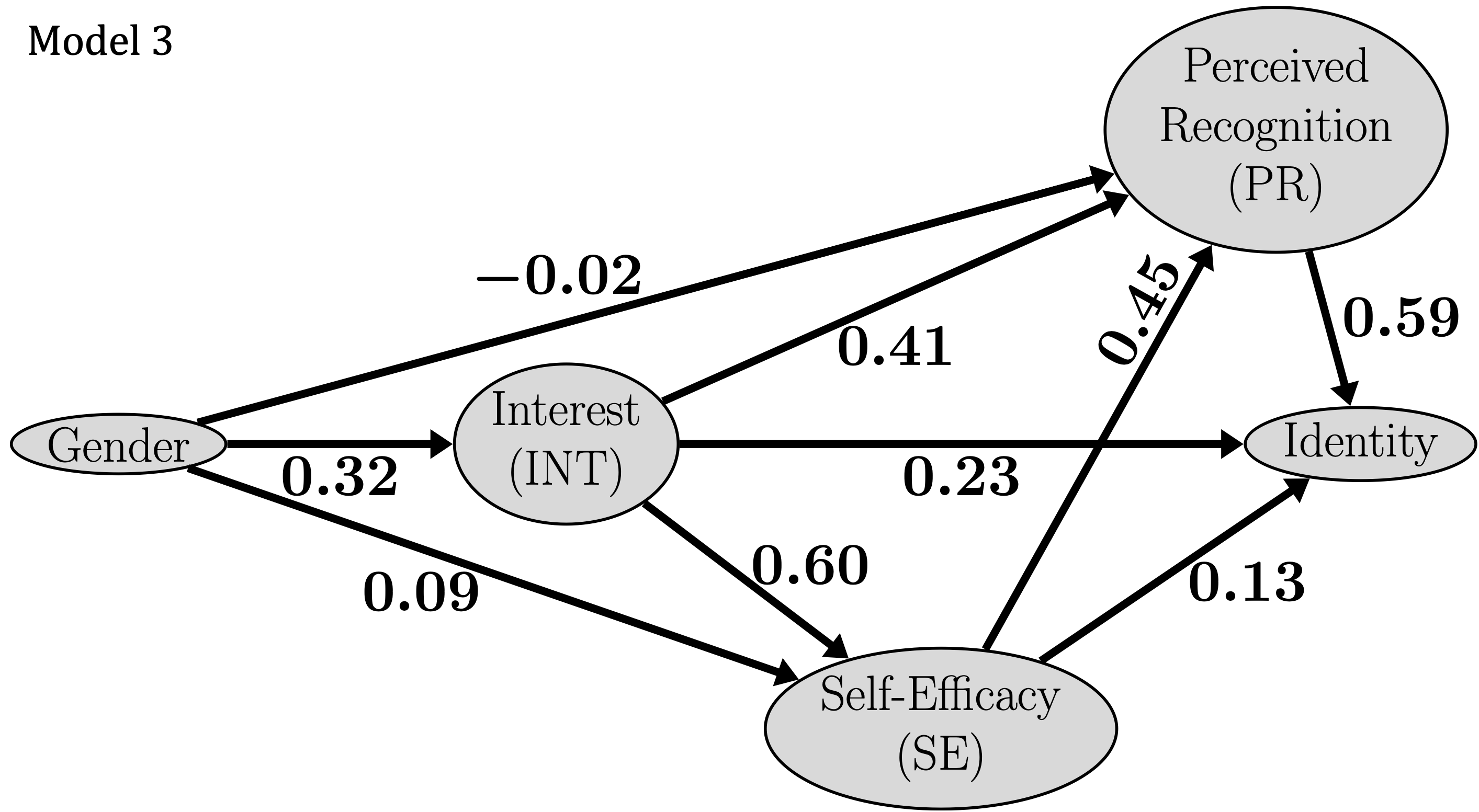}
         \caption{}
         \label{FIModel3}
     \end{subfigure}
     \hfill
     \begin{subfigure}[b]{0.45\textwidth}
         \centering
         \includegraphics[width=\textwidth]{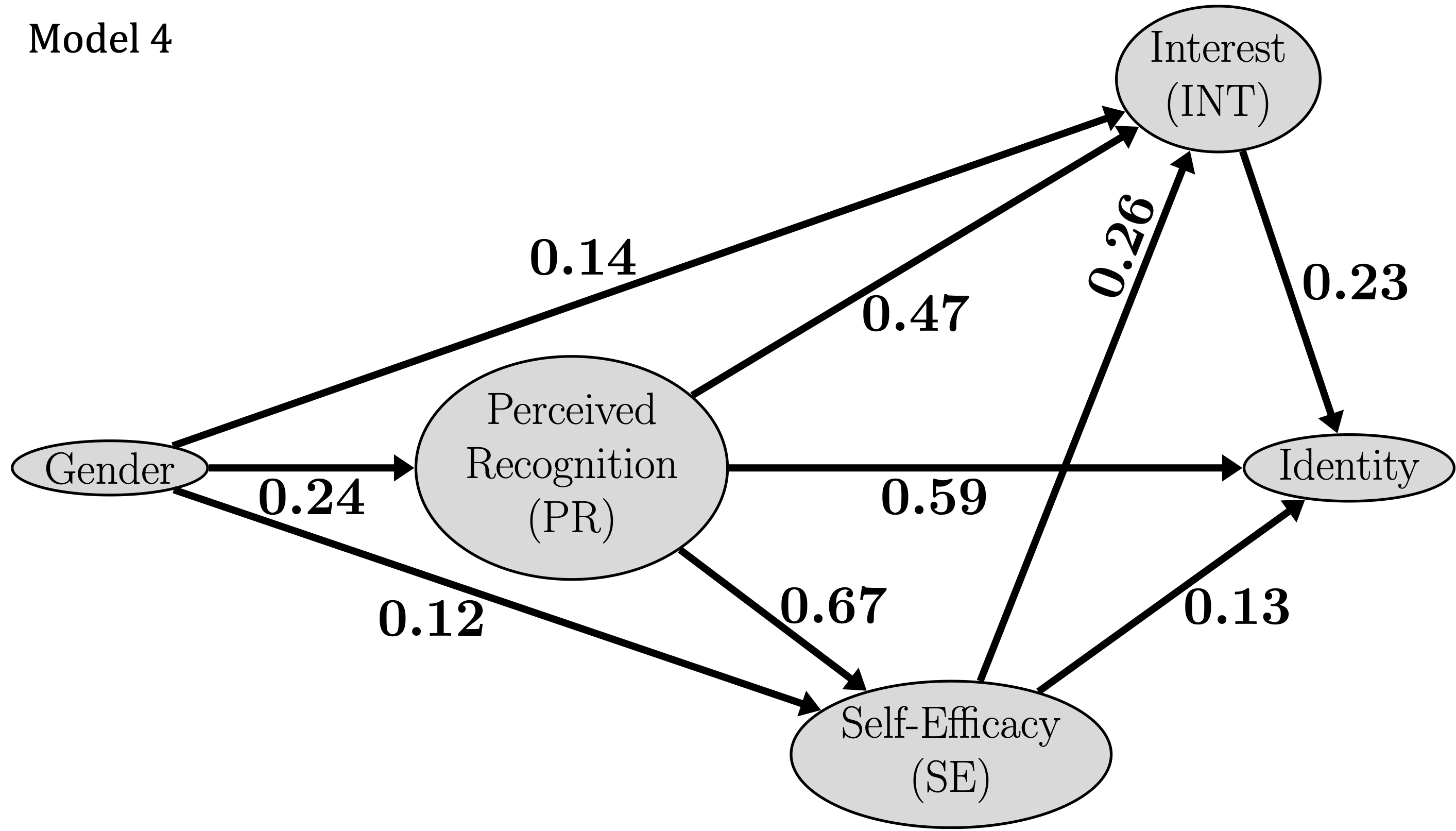}
         \caption{}
         \label{FIModel4}
     \end{subfigure}
        \caption{Models, as shown in \cite{li2023select}, represent theorized relationships between gender and physics identity mediated through three motivational variables: self-efficacy (SE), interest (INT), and perceived recognition (PR). (a) In Model 1, the three mediating variables are non-causally associated with one another. (b-d) In Models 2-4, there are causal associations between the three motivational variables: one of the three motivational variables is a common cause of the others, and another variable is a common effect of the others.}
        \label{FIModels}
\end{figure*}

To propose a concrete example of using observational data to predict the results of future interventions, we consider a recent example from Li and Singh \cite{li2023select}, who used observational data to investigate the relations between gender and four motivational constructs: self-efficacy, interest, perceived recognition, and identity.  This paper provides a good case study of these issues for the following reasons: it analyzes correlated motivational factors in a non-intervention setting; the motivational variables are highly correlated ($r > 0.6$ for all six pairwise correlations between the four motivational factors); and it deals with motivational variables which can plausibly be modeled as reciprocally co-evolving.  This paper also does the rare work of explicitly comparing alternative causal models, considering four causal models where self-efficacy, interest, and perceived recognition mediate the relationship between gender and identity.  Model 1 (Fig. \ref{FIModel1}) considers no causal association between self-efficacy, interest, and perceived recognition. Models 2-4 (Fig. \labelcref{FIModel2,FIModel3,FIModel4}) make one of these mediators a cause of the other two mediators. For instance, model 4 describes the total effect of perceived recognition on identity as the sum of a direct effect $Perceived \; Recognition \rightarrow Identity$ and indirect effects of $Perceived \; Recognition \rightarrow Interest \rightarrow Identity$ and $Perceived \; Recognition \rightarrow SE \rightarrow Identity$.  Because of the highly connected nature of these motivational constructs, all of these models could be viewed as theoretically reasonable to some degree.  For example, because it is reasonable for people to be more interested in topics that they believe they can successfully learn, $SE \rightarrow Interest$ is plausible, but because interest also likely increases engagement and learning in an area, $Interest \rightarrow SE$ is also plausible.  Similarly, although identity is an effect caused by the three other factors, identity may also be a cause that can impact one's self-efficacy, interest, or perceived recognition. Although this paper uses structural equation modeling rather than path analysis with linear regression to find the coefficients on the DAG, the conceptual issues underlying the causal analysis remain the same.

The quantitative analysis of these four models does not prove or disprove any one model over the others. Li and Singh \cite{li2023select} recognize this and give two other reasons to favor model 4 over the others: they argue that perceived recognition should be seen as a parent cause since (1) students self-report it as the causal antecedent of their later motivation in interviews and (2) perceived recognition is the factor most closely associated with instructors' actions so a model that places perceived recognition as a parent cause is the best motivator of instructor change.  However, these arguments, though reasonable, do not support the causal validity of model 4 over the others.  An individual's perceptions of a phenomenon, even regarding their own motivation, may not reflect the actual causal mechanisms of that phenomenon.  Also, having desirable implications does not make a model more causally accurate than other models. 

We propose that the best method to differentiate between these models is to compare how well they predict future interventions.  For instance, consider an intervention that aims to increase students' perceived recognition. Models 1, 2, and 3 predict that a +1 SD increase in perceived recognition should produce a +0.59 SD increase in identity and that SE and interest should not change because they are not causally associated with perceived recognition (model 1) or they are causes of perceived recognition that should not be affected when perceived recognition is directly intervened upon (models 2 and 3).  However, model 4 predicts that a +1 SD change to perceived recognition should cause a total effect of $(0.59)+(0.47)(0.23)+(0.67)(0.13+(0.26*0.23))= +0.83$ SD on identity, a +.67 SD change on SE, and a +0.47 SD change on interest.  Therefore, collecting data on the effects of intervening on perceived recognition and comparing the results against the predictions made by each theoretical model can potentially support or down-weight model 4.  Incorrect causal models conflate causal effects with non-causal associations and can misestimate how interventions on one factor will cascade through the causal system. 

In addition, the intervention should have upstream consequences as well.  Intervening on perceived recognition should also weaken or break associations with causes in the model since direct intervention will change perceived recognition so that it is no longer tied to its causes.  Importantly, these models suggest that perceived recognition is a mediator of this effect of gender on identity, so weakening the direct and indirect paths between gender and perceived recognition should weaken the total causal effect of gender and identity.  However, the intervention could have no impact on the total effect of gender on identity, which could indicate that the proposed causal model is incorrect and that perceived recognition, interest, and/or self-efficacy do not mediate the causal impact of gender on identity.  In this way, the DAGs provide an explicit, quantitative model for making quantitative predictions about the cascading effects of hypothesized interventions.

This use of observational results to motivate and predict the results of intervention studies is aligned with the ultimate goal of improving educational experiences and outcomes for students.  Debates about which theoretical model correctly describes the underlying causal relationships are only useful as far as they inspire and suggest future interventions.  Proposing (and conducting) future intervention studies motivated by these theories moves us closer to the goal of designing, testing, and disseminating instructional improvements.

\section{Summary}

``Correlation does not imply causation'' is a commonly stated aphorism that reminds researchers to err on the side of caution.  However, it is equally true that ``correlations sometimes indicate causation'' and that ``correlations can contain information about causation.'' Causal diagrams and associated rules for statistical causal inference provide a framework for extracting causal information from correlational data when appropriate (and for cautioning researchers from doing so when it is not appropriate).  While we expect that many Physics Education researchers engaged in making statistical causal inferences will be familiar with these methods, we hope that this paper helps knowledge of these techniques become more widespread in PER. This paper describes some of the well-known fundamental principles of statistical causal inference, illustrates some connections to existing PER studies, and proposes a new explicit epistemological role for observational studies as theoretical proposals for future intervention studies.  We hope that this paper provides a starting point for researchers to learn more about the causal inference methods and analysis techniques well-established outside of PER.

We close by summarizing three main takeaway points from this discussion of causal inference methods: (1)
A primary rule of causal inference is that analysis should control for confounders and not control for mediators and colliders.  Making a researcher's causal assumptions about a system explicit through a DAG provides a diagrammatic method for differentiating confounders, mediators, and colliders.  These causal inference techniques are especially important in observational studies, where observed correlations can represent both causal and non-causal associations.  The benefit of intervention studies is that direct manipulation of causal factors can break or weaken the non-causal backdoor associations opened by confounders.

(2) The biggest weakness of these causal inference techniques is that the validity of the causal inferences depends entirely on the accuracy of the proposed causal diagram.  Quantitative metrics, such as path coefficient values or goodness-of-fit statistics, can not support the causal validity of one proposed model over another.  Therefore, even seemingly reasonable causal estimates produced by quantitative analysis can be incorrect.  This highlights the importance of model selection and avoiding (explicit and implicit) claims that a causal analysis of observational data ``proves,'' ``shows,'' or ``demonstrates'' evidence for any causal claim.  Just as with theoretical research in physics, finding a quantitative result should not be taken as evidence supporting a postulated theory.

(3) We propose an explicit role for studies applying path analysis, structural equation modeling, or other analyses commonly used to draw causal inferences from observational data: motivating future intervention studies.  This role embraces the strengths of observational studies while making explicit the theoretical nature of the causal inferences drawn.  It also promotes greater coordination between observation and intervention studies to forward the science of effective instructional interventions.


\bibliography{bibliography}

\begin{thebibliography}{98}%
\makeatletter
\providecommand \@ifxundefined [1]{%
 \@ifx{#1\undefined}
}%
\providecommand \@ifnum [1]{%
 \ifnum #1\expandafter \@firstoftwo
 \else \expandafter \@secondoftwo
 \fi
}%
\providecommand \@ifx [1]{%
 \ifx #1\expandafter \@firstoftwo
 \else \expandafter \@secondoftwo
 \fi
}%
\providecommand \natexlab [1]{#1}%
\providecommand \enquote  [1]{``#1''}%
\providecommand \bibnamefont  [1]{#1}%
\providecommand \bibfnamefont [1]{#1}%
\providecommand \citenamefont [1]{#1}%
\providecommand \href@noop [0]{\@secondoftwo}%
\providecommand \href [0]{\begingroup \@sanitize@url \@href}%
\providecommand \@href[1]{\@@startlink{#1}\@@href}%
\providecommand \@@href[1]{\endgroup#1\@@endlink}%
\providecommand \@sanitize@url [0]{\catcode `\\12\catcode `\$12\catcode
  `\&12\catcode `\#12\catcode `\^12\catcode `\_12\catcode `\%12\relax}%
\providecommand \@@startlink[1]{}%
\providecommand \@@endlink[0]{}%
\providecommand \url  [0]{\begingroup\@sanitize@url \@url }%
\providecommand \@url [1]{\endgroup\@href {#1}{\urlprefix }}%
\providecommand \urlprefix  [0]{URL }%
\providecommand \Eprint [0]{\href }%
\providecommand \doibase [0]{https://doi.org/}%
\providecommand \selectlanguage [0]{\@gobble}%
\providecommand \bibinfo  [0]{\@secondoftwo}%
\providecommand \bibfield  [0]{\@secondoftwo}%
\providecommand \translation [1]{[#1]}%
\providecommand \BibitemOpen [0]{}%
\providecommand \bibitemStop [0]{}%
\providecommand \bibitemNoStop [0]{.\EOS\space}%
\providecommand \EOS [0]{\spacefactor3000\relax}%
\providecommand \BibitemShut  [1]{\csname bibitem#1\endcsname}%
\let\auto@bib@innerbib\@empty
\bibitem [{\citenamefont {Weissman}(2021)}]{weissman2021policy}%
  \BibitemOpen
  \bibfield  {author} {\bibinfo {author} {\bibfnamefont {M.~B.}\ \bibnamefont
  {Weissman}},\ }\bibfield  {title} {\bibinfo {title} {Policy recommendations
  from causal inference in physics education research},\ }\href@noop {}
  {\bibfield  {journal} {\bibinfo  {journal} {Physical Review Physics Education
  Research}\ }\textbf {\bibinfo {volume} {17}},\ \bibinfo {pages} {020118}
  (\bibinfo {year} {2021})}\BibitemShut {NoStop}%
\bibitem [{\citenamefont {Weissman}(2022)}]{weissman2022invalid}%
  \BibitemOpen
  \bibfield  {author} {\bibinfo {author} {\bibfnamefont {M.~B.}\ \bibnamefont
  {Weissman}},\ }\bibfield  {title} {\bibinfo {title} {Invalid methods and
  false answers: Physics education research and the use of gres.},\ }\href@noop
  {} {\bibfield  {journal} {\bibinfo  {journal} {Econ Journal Watch}\ }\textbf
  {\bibinfo {volume} {19}} (\bibinfo {year} {2022})}\BibitemShut {NoStop}%
\bibitem [{\citenamefont {Varian}(2016)}]{varian2016causal}%
  \BibitemOpen
  \bibfield  {author} {\bibinfo {author} {\bibfnamefont {H.~R.}\ \bibnamefont
  {Varian}},\ }\bibfield  {title} {\bibinfo {title} {Causal inference in
  economics and marketing},\ }\href@noop {} {\bibfield  {journal} {\bibinfo
  {journal} {Proceedings of the National Academy of Sciences}\ }\textbf
  {\bibinfo {volume} {113}},\ \bibinfo {pages} {7310} (\bibinfo {year}
  {2016})}\BibitemShut {NoStop}%
\bibitem [{\citenamefont {Rohrer}(2018)}]{rohrer2018thinking}%
  \BibitemOpen
  \bibfield  {author} {\bibinfo {author} {\bibfnamefont {J.~M.}\ \bibnamefont
  {Rohrer}},\ }\bibfield  {title} {\bibinfo {title} {Thinking clearly about
  correlations and causation: Graphical causal models for observational data},\
  }\href@noop {} {\bibfield  {journal} {\bibinfo  {journal} {Advances in
  methods and practices in psychological science}\ }\textbf {\bibinfo {volume}
  {1}},\ \bibinfo {pages} {27} (\bibinfo {year} {2018})}\BibitemShut {NoStop}%
\bibitem [{\citenamefont {Foster}(2010)}]{foster2010causal}%
  \BibitemOpen
  \bibfield  {author} {\bibinfo {author} {\bibfnamefont {E.~M.}\ \bibnamefont
  {Foster}},\ }\bibfield  {title} {\bibinfo {title} {Causal inference and
  developmental psychology.},\ }\href@noop {} {\bibfield  {journal} {\bibinfo
  {journal} {Developmental psychology}\ }\textbf {\bibinfo {volume} {46}},\
  \bibinfo {pages} {1454} (\bibinfo {year} {2010})}\BibitemShut {NoStop}%
\bibitem [{\citenamefont {Glass}\ \emph {et~al.}(2013)\citenamefont {Glass},
  \citenamefont {Goodman}, \citenamefont {Hern{\'a}n},\ and\ \citenamefont
  {Samet}}]{glass2013causal}%
  \BibitemOpen
  \bibfield  {author} {\bibinfo {author} {\bibfnamefont {T.~A.}\ \bibnamefont
  {Glass}}, \bibinfo {author} {\bibfnamefont {S.~N.}\ \bibnamefont {Goodman}},
  \bibinfo {author} {\bibfnamefont {M.~A.}\ \bibnamefont {Hern{\'a}n}},\ and\
  \bibinfo {author} {\bibfnamefont {J.~M.}\ \bibnamefont {Samet}},\ }\bibfield
  {title} {\bibinfo {title} {Causal inference in public health},\ }\href@noop
  {} {\bibfield  {journal} {\bibinfo  {journal} {Annual review of public
  health}\ }\textbf {\bibinfo {volume} {34}},\ \bibinfo {pages} {61} (\bibinfo
  {year} {2013})}\BibitemShut {NoStop}%
\bibitem [{\citenamefont {Gangl}(2010)}]{gangl2010causal}%
  \BibitemOpen
  \bibfield  {author} {\bibinfo {author} {\bibfnamefont {M.}~\bibnamefont
  {Gangl}},\ }\bibfield  {title} {\bibinfo {title} {Causal inference in
  sociological research},\ }\href@noop {} {\bibfield  {journal} {\bibinfo
  {journal} {Annual review of sociology}\ }\textbf {\bibinfo {volume} {36}},\
  \bibinfo {pages} {21} (\bibinfo {year} {2010})}\BibitemShut {NoStop}%
\bibitem [{\citenamefont {Keele}(2015)}]{keele2015statistics}%
  \BibitemOpen
  \bibfield  {author} {\bibinfo {author} {\bibfnamefont {L.}~\bibnamefont
  {Keele}},\ }\bibfield  {title} {\bibinfo {title} {The statistics of causal
  inference: A view from political methodology},\ }\href@noop {} {\bibfield
  {journal} {\bibinfo  {journal} {Political Analysis}\ }\textbf {\bibinfo
  {volume} {23}},\ \bibinfo {pages} {313} (\bibinfo {year} {2015})}\BibitemShut
  {NoStop}%
\bibitem [{\citenamefont {Murnane}\ and\ \citenamefont
  {Willett}(2010)}]{murnane2010methods}%
  \BibitemOpen
  \bibfield  {author} {\bibinfo {author} {\bibfnamefont {R.~J.}\ \bibnamefont
  {Murnane}}\ and\ \bibinfo {author} {\bibfnamefont {J.~B.}\ \bibnamefont
  {Willett}},\ }\href@noop {} {\emph {\bibinfo {title} {Methods matter:
  Improving causal inference in educational and social science research}}}\
  (\bibinfo  {publisher} {Oxford University Press},\ \bibinfo {year}
  {2010})\BibitemShut {NoStop}%
\bibitem [{\citenamefont {Vandenbroucke}\ \emph {et~al.}(2016)\citenamefont
  {Vandenbroucke}, \citenamefont {Broadbent},\ and\ \citenamefont
  {Pearce}}]{vandenbroucke2016causality}%
  \BibitemOpen
  \bibfield  {author} {\bibinfo {author} {\bibfnamefont {J.~P.}\ \bibnamefont
  {Vandenbroucke}}, \bibinfo {author} {\bibfnamefont {A.}~\bibnamefont
  {Broadbent}},\ and\ \bibinfo {author} {\bibfnamefont {N.}~\bibnamefont
  {Pearce}},\ }\bibfield  {title} {\bibinfo {title} {Causality and causal
  inference in epidemiology: the need for a pluralistic approach},\ }\href@noop
  {} {\bibfield  {journal} {\bibinfo  {journal} {International journal of
  epidemiology}\ }\textbf {\bibinfo {volume} {45}},\ \bibinfo {pages} {1776}
  (\bibinfo {year} {2016})}\BibitemShut {NoStop}%
\bibitem [{\citenamefont {Freedman}\ \emph {et~al.}(2007)\citenamefont
  {Freedman}, \citenamefont {Pisani},\ and\ \citenamefont
  {Purves}}]{freedmanbook}%
  \BibitemOpen
  \bibfield  {author} {\bibinfo {author} {\bibfnamefont {D.}~\bibnamefont
  {Freedman}}, \bibinfo {author} {\bibfnamefont {R.}~\bibnamefont {Pisani}},\
  and\ \bibinfo {author} {\bibfnamefont {R.}~\bibnamefont {Purves}},\
  }\href@noop {} {\emph {\bibinfo {title} {Statistics}}}\ (\bibinfo
  {publisher} {W.W. Norton \& Company},\ \bibinfo {year} {2007})\BibitemShut
  {NoStop}%
\bibitem [{\citenamefont {Imbens}(2020)}]{imbens2020potential}%
  \BibitemOpen
  \bibfield  {author} {\bibinfo {author} {\bibfnamefont {G.~W.}\ \bibnamefont
  {Imbens}},\ }\bibfield  {title} {\bibinfo {title} {Potential outcome and
  directed acyclic graph approaches to causality: Relevance for empirical
  practice in economics},\ }\href@noop {} {\bibfield  {journal} {\bibinfo
  {journal} {Journal of Economic Literature}\ }\textbf {\bibinfo {volume}
  {58}},\ \bibinfo {pages} {1129} (\bibinfo {year} {2020})}\BibitemShut
  {NoStop}%
\bibitem [{\citenamefont {Pearce}\ and\ \citenamefont
  {Lawlor}(2016)}]{pearce2016causal}%
  \BibitemOpen
  \bibfield  {author} {\bibinfo {author} {\bibfnamefont {N.}~\bibnamefont
  {Pearce}}\ and\ \bibinfo {author} {\bibfnamefont {D.~A.}\ \bibnamefont
  {Lawlor}},\ }\bibfield  {title} {\bibinfo {title} {Causal inference—so much
  more than statistics},\ }\href@noop {} {\bibfield  {journal} {\bibinfo
  {journal} {International journal of epidemiology}\ }\textbf {\bibinfo
  {volume} {45}},\ \bibinfo {pages} {1895} (\bibinfo {year}
  {2016})}\BibitemShut {NoStop}%
\bibitem [{\citenamefont {Williams}\ \emph {et~al.}(2018)\citenamefont
  {Williams}, \citenamefont {Bach}, \citenamefont {Matthiesen}, \citenamefont
  {Henriksen},\ and\ \citenamefont {Gagliardi}}]{williams2018directed}%
  \BibitemOpen
  \bibfield  {author} {\bibinfo {author} {\bibfnamefont {T.~C.}\ \bibnamefont
  {Williams}}, \bibinfo {author} {\bibfnamefont {C.~C.}\ \bibnamefont {Bach}},
  \bibinfo {author} {\bibfnamefont {N.~B.}\ \bibnamefont {Matthiesen}},
  \bibinfo {author} {\bibfnamefont {T.~B.}\ \bibnamefont {Henriksen}},\ and\
  \bibinfo {author} {\bibfnamefont {L.}~\bibnamefont {Gagliardi}},\ }\bibfield
  {title} {\bibinfo {title} {Directed acyclic graphs: a tool for causal studies
  in paediatrics},\ }\href@noop {} {\bibfield  {journal} {\bibinfo  {journal}
  {Pediatric research}\ }\textbf {\bibinfo {volume} {84}},\ \bibinfo {pages}
  {487} (\bibinfo {year} {2018})}\BibitemShut {NoStop}%
\bibitem [{\citenamefont {Imbens}\ and\ \citenamefont
  {Rubin}(2015)}]{imbens2015causal}%
  \BibitemOpen
  \bibfield  {author} {\bibinfo {author} {\bibfnamefont {G.~W.}\ \bibnamefont
  {Imbens}}\ and\ \bibinfo {author} {\bibfnamefont {D.~B.}\ \bibnamefont
  {Rubin}},\ }\href@noop {} {\emph {\bibinfo {title} {Causal inference in
  statistics, social, and biomedical sciences}}}\ (\bibinfo  {publisher}
  {Cambridge University Press},\ \bibinfo {year} {2015})\BibitemShut {NoStop}%
\bibitem [{\citenamefont {Pearl}\ and\ \citenamefont
  {Mackenzie}(2018)}]{pearl2018book}%
  \BibitemOpen
  \bibfield  {author} {\bibinfo {author} {\bibfnamefont {J.}~\bibnamefont
  {Pearl}}\ and\ \bibinfo {author} {\bibfnamefont {D.}~\bibnamefont
  {Mackenzie}},\ }\href@noop {} {\emph {\bibinfo {title} {The book of why: the
  new science of cause and effect}}}\ (\bibinfo  {publisher} {Basic books},\
  \bibinfo {year} {2018})\BibitemShut {NoStop}%
\bibitem [{\citenamefont {Greenland}\ \emph {et~al.}(1999)\citenamefont
  {Greenland}, \citenamefont {Pearl},\ and\ \citenamefont
  {Robins}}]{greenland1999causal}%
  \BibitemOpen
  \bibfield  {author} {\bibinfo {author} {\bibfnamefont {S.}~\bibnamefont
  {Greenland}}, \bibinfo {author} {\bibfnamefont {J.}~\bibnamefont {Pearl}},\
  and\ \bibinfo {author} {\bibfnamefont {J.~M.}\ \bibnamefont {Robins}},\
  }\bibfield  {title} {\bibinfo {title} {Causal diagrams for epidemiologic
  research},\ }\href@noop {} {\bibfield  {journal} {\bibinfo  {journal}
  {Epidemiology}\ ,\ \bibinfo {pages} {37}} (\bibinfo {year}
  {1999})}\BibitemShut {NoStop}%
\bibitem [{\citenamefont {Hernán}\ and\ \citenamefont
  {Robins}(2020)}]{hernan2020book}%
  \BibitemOpen
  \bibfield  {author} {\bibinfo {author} {\bibfnamefont {M.}~\bibnamefont
  {Hernán}}\ and\ \bibinfo {author} {\bibfnamefont {J.}~\bibnamefont
  {Robins}},\ }\href@noop {} {\emph {\bibinfo {title} {Causal Inference: What
  If}}}\ (\bibinfo  {publisher} {Boca Raton: Chapman \& Hall/CRC},\ \bibinfo
  {year} {2020})\BibitemShut {NoStop}%
\bibitem [{\citenamefont {Glymour}\ \emph {et~al.}(2016)\citenamefont
  {Glymour}, \citenamefont {Pearl},\ and\ \citenamefont
  {Jewell}}]{glymour2016causal}%
  \BibitemOpen
  \bibfield  {author} {\bibinfo {author} {\bibfnamefont {M.}~\bibnamefont
  {Glymour}}, \bibinfo {author} {\bibfnamefont {J.}~\bibnamefont {Pearl}},\
  and\ \bibinfo {author} {\bibfnamefont {N.~P.}\ \bibnamefont {Jewell}},\
  }\href@noop {} {\emph {\bibinfo {title} {Causal inference in statistics: A
  primer}}}\ (\bibinfo  {publisher} {John Wiley \& Sons},\ \bibinfo {year}
  {2016})\BibitemShut {NoStop}%
\bibitem [{\citenamefont {Glymour}\ \emph {et~al.}(2019)\citenamefont
  {Glymour}, \citenamefont {Zhang},\ and\ \citenamefont
  {Spirtes}}]{glymour2019review}%
  \BibitemOpen
  \bibfield  {author} {\bibinfo {author} {\bibfnamefont {C.}~\bibnamefont
  {Glymour}}, \bibinfo {author} {\bibfnamefont {K.}~\bibnamefont {Zhang}},\
  and\ \bibinfo {author} {\bibfnamefont {P.}~\bibnamefont {Spirtes}},\
  }\bibfield  {title} {\bibinfo {title} {Review of causal discovery methods
  based on graphical models},\ }\href@noop {} {\bibfield  {journal} {\bibinfo
  {journal} {Frontiers in genetics}\ }\textbf {\bibinfo {volume} {10}},\
  \bibinfo {pages} {524} (\bibinfo {year} {2019})}\BibitemShut {NoStop}%
\bibitem [{\citenamefont {Morgan}(2013)}]{morgan2013handbook}%
  \BibitemOpen
  \bibfield  {author} {\bibinfo {author} {\bibfnamefont {S.~L.}\ \bibnamefont
  {Morgan}},\ }\href@noop {} {\emph {\bibinfo {title} {Handbook of causal
  analysis for social research}}}\ (\bibinfo  {publisher} {Springer},\ \bibinfo
  {year} {2013})\BibitemShut {NoStop}%
\bibitem [{\citenamefont {Peters}\ \emph {et~al.}(2017)\citenamefont {Peters},
  \citenamefont {Janzing},\ and\ \citenamefont
  {Sch{\"o}lkopf}}]{peters2017elements}%
  \BibitemOpen
  \bibfield  {author} {\bibinfo {author} {\bibfnamefont {J.}~\bibnamefont
  {Peters}}, \bibinfo {author} {\bibfnamefont {D.}~\bibnamefont {Janzing}},\
  and\ \bibinfo {author} {\bibfnamefont {B.}~\bibnamefont {Sch{\"o}lkopf}},\
  }\href@noop {} {\emph {\bibinfo {title} {Elements of causal inference:
  foundations and learning algorithms}}}\ (\bibinfo  {publisher} {The MIT
  Press},\ \bibinfo {year} {2017})\BibitemShut {NoStop}%
\bibitem [{\citenamefont {Morgan}\ and\ \citenamefont
  {Winship}(2015)}]{morgan2015counterfactuals}%
  \BibitemOpen
  \bibfield  {author} {\bibinfo {author} {\bibfnamefont {S.~L.}\ \bibnamefont
  {Morgan}}\ and\ \bibinfo {author} {\bibfnamefont {C.}~\bibnamefont
  {Winship}},\ }\href@noop {} {\emph {\bibinfo {title} {Counterfactuals and
  causal inference}}}\ (\bibinfo  {publisher} {Cambridge University Press},\
  \bibinfo {year} {2015})\BibitemShut {NoStop}%
\bibitem [{\citenamefont {Pearl}(2013)}]{pearl2013linear}%
  \BibitemOpen
  \bibfield  {author} {\bibinfo {author} {\bibfnamefont {J.}~\bibnamefont
  {Pearl}},\ }\bibfield  {title} {\bibinfo {title} {Linear models: A useful
  “microscope” for causal analysis},\ }\href@noop {} {\bibfield  {journal}
  {\bibinfo  {journal} {Journal of Causal Inference}\ }\textbf {\bibinfo
  {volume} {1}},\ \bibinfo {pages} {155} (\bibinfo {year} {2013})}\BibitemShut
  {NoStop}%
\bibitem [{\citenamefont {Wright}(1921)}]{wright1921correlation}%
  \BibitemOpen
  \bibfield  {author} {\bibinfo {author} {\bibfnamefont {S.}~\bibnamefont
  {Wright}},\ }\bibfield  {title} {\bibinfo {title} {Correlation and
  causation},\ }\href@noop {} {\bibfield  {journal} {\bibinfo  {journal}
  {Journal of Agricultural Research}\ }\textbf {\bibinfo {volume} {20}},\
  \bibinfo {pages} {557} (\bibinfo {year} {1921})}\BibitemShut {NoStop}%
\bibitem [{\citenamefont {Wright}(1918)}]{wright1918nature}%
  \BibitemOpen
  \bibfield  {author} {\bibinfo {author} {\bibfnamefont {S.}~\bibnamefont
  {Wright}},\ }\bibfield  {title} {\bibinfo {title} {On the nature of size
  factors},\ }\href@noop {} {\bibfield  {journal} {\bibinfo  {journal}
  {Genetics}\ }\textbf {\bibinfo {volume} {3}},\ \bibinfo {pages} {367}
  (\bibinfo {year} {1918})}\BibitemShut {NoStop}%
\bibitem [{\citenamefont {Wright}(1934)}]{wright1934method}%
  \BibitemOpen
  \bibfield  {author} {\bibinfo {author} {\bibfnamefont {S.}~\bibnamefont
  {Wright}},\ }\bibfield  {title} {\bibinfo {title} {The method of path
  coefficients},\ }\href@noop {} {\bibfield  {journal} {\bibinfo  {journal}
  {The annals of mathematical statistics}\ }\textbf {\bibinfo {volume} {5}},\
  \bibinfo {pages} {161} (\bibinfo {year} {1934})}\BibitemShut {NoStop}%
\bibitem [{\citenamefont {Saris}\ and\ \citenamefont
  {Stronkhorst}(1983)}]{saris1983introduction}%
  \BibitemOpen
  \bibfield  {author} {\bibinfo {author} {\bibfnamefont {W.~E.}\ \bibnamefont
  {Saris}}\ and\ \bibinfo {author} {\bibfnamefont {L.~H.}\ \bibnamefont
  {Stronkhorst}},\ }\href@noop {} {\emph {\bibinfo {title} {Introduction to
  Causal Modeling in Nonexperimental Research: The LISREL Approach}}}\
  (\bibinfo  {publisher} {Sociometric Research Foundation},\ \bibinfo {year}
  {1983})\BibitemShut {NoStop}%
\bibitem [{\citenamefont {Shmueli}(2010)}]{shmueli2010explain}%
  \BibitemOpen
  \bibfield  {author} {\bibinfo {author} {\bibfnamefont {G.}~\bibnamefont
  {Shmueli}},\ }\bibfield  {title} {\bibinfo {title} {To explain or to
  predict?},\ }\href@noop {} {\bibfield  {journal} {\bibinfo  {journal}
  {Statistical Science}\ }\textbf {\bibinfo {volume} {25}},\ \bibinfo {pages}
  {289} (\bibinfo {year} {2010})}\BibitemShut {NoStop}%
\bibitem [{\citenamefont {Kuhn}\ \emph {et~al.}(2013)\citenamefont {Kuhn},
  \citenamefont {Johnson} \emph {et~al.}}]{kuhn2013applied}%
  \BibitemOpen
  \bibfield  {author} {\bibinfo {author} {\bibfnamefont {M.}~\bibnamefont
  {Kuhn}}, \bibinfo {author} {\bibfnamefont {K.}~\bibnamefont {Johnson}}, \emph
  {et~al.},\ }\href@noop {} {\emph {\bibinfo {title} {Applied predictive
  modeling}}},\ Vol.~\bibinfo {volume} {26}\ (\bibinfo  {publisher}
  {Springer},\ \bibinfo {year} {2013})\BibitemShut {NoStop}%
\bibitem [{\citenamefont {Burkholder}\ \emph {et~al.}(2021)\citenamefont
  {Burkholder}, \citenamefont {Murillo-Gonzalez},\ and\ \citenamefont
  {Wieman}}]{burkholder2021importance}%
  \BibitemOpen
  \bibfield  {author} {\bibinfo {author} {\bibfnamefont {E.~W.}\ \bibnamefont
  {Burkholder}}, \bibinfo {author} {\bibfnamefont {G.}~\bibnamefont
  {Murillo-Gonzalez}},\ and\ \bibinfo {author} {\bibfnamefont {C.}~\bibnamefont
  {Wieman}},\ }\bibfield  {title} {\bibinfo {title} {Importance of math
  prerequisites for performance in introductory physics},\ }\href@noop {}
  {\bibfield  {journal} {\bibinfo  {journal} {Physical Review Physics Education
  Research}\ }\textbf {\bibinfo {volume} {17}},\ \bibinfo {pages} {010108}
  (\bibinfo {year} {2021})}\BibitemShut {NoStop}%
\bibitem [{\citenamefont {McCammon}\ \emph {et~al.}(1988)\citenamefont
  {McCammon}, \citenamefont {Golden},\ and\ \citenamefont
  {Wuensch}}]{mccammon1988predicting}%
  \BibitemOpen
  \bibfield  {author} {\bibinfo {author} {\bibfnamefont {S.}~\bibnamefont
  {McCammon}}, \bibinfo {author} {\bibfnamefont {J.}~\bibnamefont {Golden}},\
  and\ \bibinfo {author} {\bibfnamefont {K.~L.}\ \bibnamefont {Wuensch}},\
  }\bibfield  {title} {\bibinfo {title} {Predicting course performance in
  freshman and sophomore physics courses: Women are more predictable than
  men},\ }\href@noop {} {\bibfield  {journal} {\bibinfo  {journal} {Journal of
  Research in Science Teaching}\ }\textbf {\bibinfo {volume} {25}},\ \bibinfo
  {pages} {501} (\bibinfo {year} {1988})}\BibitemShut {NoStop}%
\bibitem [{\citenamefont {Verostek}\ \emph {et~al.}(2021)\citenamefont
  {Verostek}, \citenamefont {Miller},\ and\ \citenamefont
  {Zwickl}}]{verostek2021analyzing}%
  \BibitemOpen
  \bibfield  {author} {\bibinfo {author} {\bibfnamefont {M.}~\bibnamefont
  {Verostek}}, \bibinfo {author} {\bibfnamefont {C.~W.}\ \bibnamefont
  {Miller}},\ and\ \bibinfo {author} {\bibfnamefont {B.}~\bibnamefont
  {Zwickl}},\ }\bibfield  {title} {\bibinfo {title} {Analyzing admissions
  metrics as predictors of graduate gpa and whether graduate gpa mediates ph.
  d. completion},\ }\href@noop {} {\bibfield  {journal} {\bibinfo  {journal}
  {Physical Review Physics Education Research}\ }\textbf {\bibinfo {volume}
  {17}},\ \bibinfo {pages} {020115} (\bibinfo {year} {2021})}\BibitemShut
  {NoStop}%
\bibitem [{\citenamefont {Cwik}\ and\ \citenamefont
  {Singh}(2022)}]{cwik2022students}%
  \BibitemOpen
  \bibfield  {author} {\bibinfo {author} {\bibfnamefont {S.}~\bibnamefont
  {Cwik}}\ and\ \bibinfo {author} {\bibfnamefont {C.}~\bibnamefont {Singh}},\
  }\bibfield  {title} {\bibinfo {title} {Students’ sense of belonging in
  introductory physics course for bioscience majors predicts their grade},\
  }\href@noop {} {\bibfield  {journal} {\bibinfo  {journal} {Physical Review
  Physics Education Research}\ }\textbf {\bibinfo {volume} {18}},\ \bibinfo
  {pages} {010139} (\bibinfo {year} {2022})}\BibitemShut {NoStop}%
\bibitem [{\citenamefont {Salehi}\ \emph {et~al.}(2019)\citenamefont {Salehi},
  \citenamefont {Burkholder}, \citenamefont {Lepage}, \citenamefont {Pollock},\
  and\ \citenamefont {Wieman}}]{salehi2019demographic}%
  \BibitemOpen
  \bibfield  {author} {\bibinfo {author} {\bibfnamefont {S.}~\bibnamefont
  {Salehi}}, \bibinfo {author} {\bibfnamefont {E.}~\bibnamefont {Burkholder}},
  \bibinfo {author} {\bibfnamefont {G.~P.}\ \bibnamefont {Lepage}}, \bibinfo
  {author} {\bibfnamefont {S.}~\bibnamefont {Pollock}},\ and\ \bibinfo {author}
  {\bibfnamefont {C.}~\bibnamefont {Wieman}},\ }\bibfield  {title} {\bibinfo
  {title} {Demographic gaps or preparation gaps?: The large impact of incoming
  preparation on performance of students in introductory physics},\ }\href@noop
  {} {\bibfield  {journal} {\bibinfo  {journal} {Physical Review Physics
  Education Research}\ }\textbf {\bibinfo {volume} {15}},\ \bibinfo {pages}
  {020114} (\bibinfo {year} {2019})}\BibitemShut {NoStop}%
\bibitem [{\citenamefont {Huang}\ and\ \citenamefont
  {Fang}(2013)}]{huang2013predicting}%
  \BibitemOpen
  \bibfield  {author} {\bibinfo {author} {\bibfnamefont {S.}~\bibnamefont
  {Huang}}\ and\ \bibinfo {author} {\bibfnamefont {N.}~\bibnamefont {Fang}},\
  }\bibfield  {title} {\bibinfo {title} {Predicting student academic
  performance in an engineering dynamics course: A comparison of four types of
  predictive mathematical models},\ }\href@noop {} {\bibfield  {journal}
  {\bibinfo  {journal} {Computers \& Education}\ }\textbf {\bibinfo {volume}
  {61}},\ \bibinfo {pages} {133} (\bibinfo {year} {2013})}\BibitemShut
  {NoStop}%
\bibitem [{\citenamefont {Ting}\ and\ \citenamefont
  {Man}(2001)}]{ting2001predicting}%
  \BibitemOpen
  \bibfield  {author} {\bibinfo {author} {\bibfnamefont {S.-M.~R.}\
  \bibnamefont {Ting}}\ and\ \bibinfo {author} {\bibfnamefont {R.}~\bibnamefont
  {Man}},\ }\bibfield  {title} {\bibinfo {title} {Predicting academic success
  of first-year engineering students from standardized test scores and
  psychosocial variables},\ }\href@noop {} {\bibfield  {journal} {\bibinfo
  {journal} {International Journal of Engineering Education}\ }\textbf
  {\bibinfo {volume} {17}},\ \bibinfo {pages} {75} (\bibinfo {year}
  {2001})}\BibitemShut {NoStop}%
\bibitem [{\citenamefont {Ransdell}(2001)}]{ransdell2001predicting}%
  \BibitemOpen
  \bibfield  {author} {\bibinfo {author} {\bibfnamefont {S.}~\bibnamefont
  {Ransdell}},\ }\bibfield  {title} {\bibinfo {title} {Predicting college
  success: The importance of ability and non-cognitive variables},\ }\href@noop
  {} {\bibfield  {journal} {\bibinfo  {journal} {International journal of
  educational Research}\ }\textbf {\bibinfo {volume} {35}},\ \bibinfo {pages}
  {357} (\bibinfo {year} {2001})}\BibitemShut {NoStop}%
\bibitem [{\citenamefont {Pokay}\ and\ \citenamefont
  {Blumenfeld}(1990)}]{pokay1990predicting}%
  \BibitemOpen
  \bibfield  {author} {\bibinfo {author} {\bibfnamefont {P.}~\bibnamefont
  {Pokay}}\ and\ \bibinfo {author} {\bibfnamefont {P.~C.}\ \bibnamefont
  {Blumenfeld}},\ }\bibfield  {title} {\bibinfo {title} {Predicting achievement
  early and late in the semester: The role of motivation and use of learning
  strategies.},\ }\href@noop {} {\bibfield  {journal} {\bibinfo  {journal}
  {Journal of educational psychology}\ }\textbf {\bibinfo {volume} {82}},\
  \bibinfo {pages} {41} (\bibinfo {year} {1990})}\BibitemShut {NoStop}%
\bibitem [{\citenamefont {Cohen}\ \emph {et~al.}(2017)\citenamefont {Cohen},
  \citenamefont {Manion},\ and\ \citenamefont {Morrison}}]{cohen2017research}%
  \BibitemOpen
  \bibfield  {author} {\bibinfo {author} {\bibfnamefont {L.}~\bibnamefont
  {Cohen}}, \bibinfo {author} {\bibfnamefont {L.}~\bibnamefont {Manion}},\ and\
  \bibinfo {author} {\bibfnamefont {K.}~\bibnamefont {Morrison}},\ }\href@noop
  {} {\emph {\bibinfo {title} {Research methods in education}}}\ (\bibinfo
  {publisher} {routledge},\ \bibinfo {year} {2017})\BibitemShut {NoStop}%
\bibitem [{\citenamefont {Zabriskie}\ \emph {et~al.}(2019)\citenamefont
  {Zabriskie}, \citenamefont {Yang}, \citenamefont {DeVore},\ and\
  \citenamefont {Stewart}}]{zabriskie2019using}%
  \BibitemOpen
  \bibfield  {author} {\bibinfo {author} {\bibfnamefont {C.}~\bibnamefont
  {Zabriskie}}, \bibinfo {author} {\bibfnamefont {J.}~\bibnamefont {Yang}},
  \bibinfo {author} {\bibfnamefont {S.}~\bibnamefont {DeVore}},\ and\ \bibinfo
  {author} {\bibfnamefont {J.}~\bibnamefont {Stewart}},\ }\bibfield  {title}
  {\bibinfo {title} {Using machine learning to predict physics course
  outcomes},\ }\href@noop {} {\bibfield  {journal} {\bibinfo  {journal}
  {Physical Review Physics Education Research}\ }\textbf {\bibinfo {volume}
  {15}},\ \bibinfo {pages} {020120} (\bibinfo {year} {2019})}\BibitemShut
  {NoStop}%
\bibitem [{\citenamefont {Sadler}\ and\ \citenamefont
  {Tai}(2001)}]{sadler2001success}%
  \BibitemOpen
  \bibfield  {author} {\bibinfo {author} {\bibfnamefont {P.~M.}\ \bibnamefont
  {Sadler}}\ and\ \bibinfo {author} {\bibfnamefont {R.~H.}\ \bibnamefont
  {Tai}},\ }\bibfield  {title} {\bibinfo {title} {Success in introductory
  college physics: The role of high school preparation},\ }\href@noop {}
  {\bibfield  {journal} {\bibinfo  {journal} {Science Education}\ }\textbf
  {\bibinfo {volume} {85}},\ \bibinfo {pages} {111} (\bibinfo {year}
  {2001})}\BibitemShut {NoStop}%
\bibitem [{\citenamefont {Hazari}\ \emph {et~al.}(2007)\citenamefont {Hazari},
  \citenamefont {Tai},\ and\ \citenamefont {Sadler}}]{hazari2007gender}%
  \BibitemOpen
  \bibfield  {author} {\bibinfo {author} {\bibfnamefont {Z.}~\bibnamefont
  {Hazari}}, \bibinfo {author} {\bibfnamefont {R.~H.}\ \bibnamefont {Tai}},\
  and\ \bibinfo {author} {\bibfnamefont {P.~M.}\ \bibnamefont {Sadler}},\
  }\bibfield  {title} {\bibinfo {title} {Gender differences in introductory
  university physics performance: The influence of high school physics
  preparation and affective factors},\ }\href@noop {} {\bibfield  {journal}
  {\bibinfo  {journal} {Science education}\ }\textbf {\bibinfo {volume} {91}},\
  \bibinfo {pages} {847} (\bibinfo {year} {2007})}\BibitemShut {NoStop}%
\bibitem [{\citenamefont {Kitsantas}\ \emph {et~al.}(2008)\citenamefont
  {Kitsantas}, \citenamefont {Winsler},\ and\ \citenamefont
  {Huie}}]{kitsantas2008self}%
  \BibitemOpen
  \bibfield  {author} {\bibinfo {author} {\bibfnamefont {A.}~\bibnamefont
  {Kitsantas}}, \bibinfo {author} {\bibfnamefont {A.}~\bibnamefont {Winsler}},\
  and\ \bibinfo {author} {\bibfnamefont {F.}~\bibnamefont {Huie}},\ }\bibfield
  {title} {\bibinfo {title} {Self-regulation and ability predictors of academic
  success during college: A predictive validity study},\ }\href@noop {}
  {\bibfield  {journal} {\bibinfo  {journal} {Journal of advanced academics}\
  }\textbf {\bibinfo {volume} {20}},\ \bibinfo {pages} {42} (\bibinfo {year}
  {2008})}\BibitemShut {NoStop}%
\bibitem [{\citenamefont {Rimfeld}\ \emph {et~al.}(2016)\citenamefont
  {Rimfeld}, \citenamefont {Kovas}, \citenamefont {Dale},\ and\ \citenamefont
  {Plomin}}]{rimfeld2016true}%
  \BibitemOpen
  \bibfield  {author} {\bibinfo {author} {\bibfnamefont {K.}~\bibnamefont
  {Rimfeld}}, \bibinfo {author} {\bibfnamefont {Y.}~\bibnamefont {Kovas}},
  \bibinfo {author} {\bibfnamefont {P.~S.}\ \bibnamefont {Dale}},\ and\
  \bibinfo {author} {\bibfnamefont {R.}~\bibnamefont {Plomin}},\ }\bibfield
  {title} {\bibinfo {title} {True grit and genetics: Predicting academic
  achievement from personality.},\ }\href@noop {} {\bibfield  {journal}
  {\bibinfo  {journal} {Journal of personality and social psychology}\ }\textbf
  {\bibinfo {volume} {111}},\ \bibinfo {pages} {780} (\bibinfo {year}
  {2016})}\BibitemShut {NoStop}%
\bibitem [{\citenamefont {Pearl}(1995)}]{pearl1995causal}%
  \BibitemOpen
  \bibfield  {author} {\bibinfo {author} {\bibfnamefont {J.}~\bibnamefont
  {Pearl}},\ }\bibfield  {title} {\bibinfo {title} {Causal diagrams for
  empirical research},\ }\href@noop {} {\bibfield  {journal} {\bibinfo
  {journal} {Biometrika}\ }\textbf {\bibinfo {volume} {82}},\ \bibinfo {pages}
  {669} (\bibinfo {year} {1995})}\BibitemShut {NoStop}%
\bibitem [{\citenamefont {Spirtes}\ \emph {et~al.}(2000)\citenamefont
  {Spirtes}, \citenamefont {Glymour},\ and\ \citenamefont
  {Scheines}}]{spirtes2000causation}%
  \BibitemOpen
  \bibfield  {author} {\bibinfo {author} {\bibfnamefont {P.}~\bibnamefont
  {Spirtes}}, \bibinfo {author} {\bibfnamefont {C.~N.}\ \bibnamefont
  {Glymour}},\ and\ \bibinfo {author} {\bibfnamefont {R.}~\bibnamefont
  {Scheines}},\ }\href@noop {} {\emph {\bibinfo {title} {Causation, prediction,
  and search}}}\ (\bibinfo  {publisher} {MIT press},\ \bibinfo {year}
  {2000})\BibitemShut {NoStop}%
\bibitem [{\citenamefont {{R Core Team}}(2022)}]{Rstudio}%
  \BibitemOpen
  \bibfield  {author} {\bibinfo {author} {\bibnamefont {{R Core Team}}},\
  }\href {https://www.R-project.org/} {\emph {\bibinfo {title} {R: A Language
  and Environment for Statistical Computing}}},\ \bibinfo {organization} {R
  Foundation for Statistical Computing},\ \bibinfo {address} {Vienna, Austria}
  (\bibinfo {year} {2022})\BibitemShut {NoStop}%
\bibitem [{\citenamefont {Gelman}\ and\ \citenamefont
  {Hill}(2006)}]{gelman2006data}%
  \BibitemOpen
  \bibfield  {author} {\bibinfo {author} {\bibfnamefont {A.}~\bibnamefont
  {Gelman}}\ and\ \bibinfo {author} {\bibfnamefont {J.}~\bibnamefont {Hill}},\
  }\href@noop {} {\emph {\bibinfo {title} {Data analysis using regression and
  multilevel/hierarchical models}}}\ (\bibinfo  {publisher} {Cambridge
  university press},\ \bibinfo {year} {2006})\BibitemShut {NoStop}%
\bibitem [{\citenamefont {Clarke}(2005)}]{clarke2005phantom}%
  \BibitemOpen
  \bibfield  {author} {\bibinfo {author} {\bibfnamefont {K.~A.}\ \bibnamefont
  {Clarke}},\ }\bibfield  {title} {\bibinfo {title} {The phantom menace:
  Omitted variable bias in econometric research},\ }\href@noop {} {\bibfield
  {journal} {\bibinfo  {journal} {Conflict management and peace science}\
  }\textbf {\bibinfo {volume} {22}},\ \bibinfo {pages} {341} (\bibinfo {year}
  {2005})}\BibitemShut {NoStop}%
\bibitem [{\citenamefont {Clarke}(2009)}]{clarke2009return}%
  \BibitemOpen
  \bibfield  {author} {\bibinfo {author} {\bibfnamefont {K.~A.}\ \bibnamefont
  {Clarke}},\ }\bibfield  {title} {\bibinfo {title} {Return of the phantom
  menace: Omitted variable bias in political research},\ }\href@noop {}
  {\bibfield  {journal} {\bibinfo  {journal} {Conflict Management and Peace
  Science}\ }\textbf {\bibinfo {volume} {26}},\ \bibinfo {pages} {46} (\bibinfo
  {year} {2009})}\BibitemShut {NoStop}%
\bibitem [{\citenamefont {Riegg}(2008)}]{riegg2008causal}%
  \BibitemOpen
  \bibfield  {author} {\bibinfo {author} {\bibfnamefont {S.~K.}\ \bibnamefont
  {Riegg}},\ }\bibfield  {title} {\bibinfo {title} {Causal inference and
  omitted variable bias in financial aid research: Assessing solutions},\
  }\href@noop {} {\bibfield  {journal} {\bibinfo  {journal} {The Review of
  Higher Education}\ }\textbf {\bibinfo {volume} {31}},\ \bibinfo {pages} {329}
  (\bibinfo {year} {2008})}\BibitemShut {NoStop}%
\bibitem [{\citenamefont {Walsh}\ \emph {et~al.}(2021)\citenamefont {Walsh},
  \citenamefont {Stein}, \citenamefont {Tapping}, \citenamefont {Smith},\ and\
  \citenamefont {Holmes}}]{walsh2021exploring}%
  \BibitemOpen
  \bibfield  {author} {\bibinfo {author} {\bibfnamefont {C.}~\bibnamefont
  {Walsh}}, \bibinfo {author} {\bibfnamefont {M.~M.}\ \bibnamefont {Stein}},
  \bibinfo {author} {\bibfnamefont {R.}~\bibnamefont {Tapping}}, \bibinfo
  {author} {\bibfnamefont {E.~M.}\ \bibnamefont {Smith}},\ and\ \bibinfo
  {author} {\bibfnamefont {N.}~\bibnamefont {Holmes}},\ }\bibfield  {title}
  {\bibinfo {title} {Exploring the effects of omitted variable bias in physics
  education research},\ }\href@noop {} {\bibfield  {journal} {\bibinfo
  {journal} {Physical Review Physics Education Research}\ }\textbf {\bibinfo
  {volume} {17}},\ \bibinfo {pages} {010119} (\bibinfo {year}
  {2021})}\BibitemShut {NoStop}%
\bibitem [{\citenamefont {Hutchison}\ and\ \citenamefont
  {Styles}(2010)}]{hutchison2010guide}%
  \BibitemOpen
  \bibfield  {author} {\bibinfo {author} {\bibfnamefont {D.}~\bibnamefont
  {Hutchison}}\ and\ \bibinfo {author} {\bibfnamefont {B.}~\bibnamefont
  {Styles}},\ }\href@noop {} {\emph {\bibinfo {title} {A guide to running
  randomised controlled trials for educational researchers}}}\ (\bibinfo
  {publisher} {NFER Slough},\ \bibinfo {year} {2010})\BibitemShut {NoStop}%
\bibitem [{\citenamefont {Torgerson}\ \emph {et~al.}(2013)\citenamefont
  {Torgerson}, \citenamefont {Wiggins}, \citenamefont {Torgerson},
  \citenamefont {Ainsworth},\ and\ \citenamefont
  {Hewitt}}]{torgerson2013every}%
  \BibitemOpen
  \bibfield  {author} {\bibinfo {author} {\bibfnamefont {C.}~\bibnamefont
  {Torgerson}}, \bibinfo {author} {\bibfnamefont {A.}~\bibnamefont {Wiggins}},
  \bibinfo {author} {\bibfnamefont {D.}~\bibnamefont {Torgerson}}, \bibinfo
  {author} {\bibfnamefont {H.}~\bibnamefont {Ainsworth}},\ and\ \bibinfo
  {author} {\bibfnamefont {C.}~\bibnamefont {Hewitt}},\ }\bibfield  {title}
  {\bibinfo {title} {Every child counts: Testing policy effectiveness using a
  randomised controlled trial, designed, conducted and reported to consort
  standards},\ }\href@noop {} {\bibfield  {journal} {\bibinfo  {journal}
  {Research in Mathematics Education}\ }\textbf {\bibinfo {volume} {15}},\
  \bibinfo {pages} {141} (\bibinfo {year} {2013})}\BibitemShut {NoStop}%
\bibitem [{\citenamefont {Hedges}\ and\ \citenamefont
  {Schauer}(2018)}]{hedges2018randomised}%
  \BibitemOpen
  \bibfield  {author} {\bibinfo {author} {\bibfnamefont {L.~V.}\ \bibnamefont
  {Hedges}}\ and\ \bibinfo {author} {\bibfnamefont {J.}~\bibnamefont
  {Schauer}},\ }\bibfield  {title} {\bibinfo {title} {Randomised trials in
  education in the usa},\ }\href@noop {} {\bibfield  {journal} {\bibinfo
  {journal} {Educational research}\ }\textbf {\bibinfo {volume} {60}},\
  \bibinfo {pages} {265} (\bibinfo {year} {2018})}\BibitemShut {NoStop}%
\bibitem [{\citenamefont {Greenland}(2003)}]{greenland2003quantifying}%
  \BibitemOpen
  \bibfield  {author} {\bibinfo {author} {\bibfnamefont {S.}~\bibnamefont
  {Greenland}},\ }\bibfield  {title} {\bibinfo {title} {Quantifying biases in
  causal models: classical confounding vs collider-stratification bias},\
  }\href@noop {} {\bibfield  {journal} {\bibinfo  {journal} {Epidemiology}\
  }\textbf {\bibinfo {volume} {14}},\ \bibinfo {pages} {300} (\bibinfo {year}
  {2003})}\BibitemShut {NoStop}%
\bibitem [{\citenamefont {Whitcomb}\ \emph {et~al.}(2009)\citenamefont
  {Whitcomb}, \citenamefont {Schisterman}, \citenamefont {Perkins},\ and\
  \citenamefont {Platt}}]{whitcomb2009quantification}%
  \BibitemOpen
  \bibfield  {author} {\bibinfo {author} {\bibfnamefont {B.~W.}\ \bibnamefont
  {Whitcomb}}, \bibinfo {author} {\bibfnamefont {E.~F.}\ \bibnamefont
  {Schisterman}}, \bibinfo {author} {\bibfnamefont {N.~J.}\ \bibnamefont
  {Perkins}},\ and\ \bibinfo {author} {\bibfnamefont {R.~W.}\ \bibnamefont
  {Platt}},\ }\bibfield  {title} {\bibinfo {title} {Quantification of
  collider-stratification bias and the birthweight paradox},\ }\href@noop {}
  {\bibfield  {journal} {\bibinfo  {journal} {Paediatric and perinatal
  epidemiology}\ }\textbf {\bibinfo {volume} {23}},\ \bibinfo {pages} {394}
  (\bibinfo {year} {2009})}\BibitemShut {NoStop}%
\bibitem [{\citenamefont {Elwert}\ and\ \citenamefont
  {Winship}(2014)}]{elwert2014endogenous}%
  \BibitemOpen
  \bibfield  {author} {\bibinfo {author} {\bibfnamefont {F.}~\bibnamefont
  {Elwert}}\ and\ \bibinfo {author} {\bibfnamefont {C.}~\bibnamefont
  {Winship}},\ }\bibfield  {title} {\bibinfo {title} {Endogenous selection
  bias: The problem of conditioning on a collider variable},\ }\href@noop {}
  {\bibfield  {journal} {\bibinfo  {journal} {Annual review of sociology}\
  }\textbf {\bibinfo {volume} {40}},\ \bibinfo {pages} {31} (\bibinfo {year}
  {2014})}\BibitemShut {NoStop}%
\bibitem [{\citenamefont {Banack}\ and\ \citenamefont
  {Kaufman}(2015)}]{banack2015bad}%
  \BibitemOpen
  \bibfield  {author} {\bibinfo {author} {\bibfnamefont {H.~R.}\ \bibnamefont
  {Banack}}\ and\ \bibinfo {author} {\bibfnamefont {J.~S.}\ \bibnamefont
  {Kaufman}},\ }\bibfield  {title} {\bibinfo {title} {From bad to worse:
  collider stratification amplifies confounding bias in the “obesity
  paradox”},\ }\href@noop {} {\bibfield  {journal} {\bibinfo  {journal}
  {European journal of epidemiology}\ }\textbf {\bibinfo {volume} {30}},\
  \bibinfo {pages} {1111} (\bibinfo {year} {2015})}\BibitemShut {NoStop}%
\bibitem [{\citenamefont {Sperrin}\ \emph {et~al.}(2016)\citenamefont
  {Sperrin}, \citenamefont {Candlish}, \citenamefont {Badrick}, \citenamefont
  {Renehan},\ and\ \citenamefont {Buchan}}]{sperrin2016collider}%
  \BibitemOpen
  \bibfield  {author} {\bibinfo {author} {\bibfnamefont {M.}~\bibnamefont
  {Sperrin}}, \bibinfo {author} {\bibfnamefont {J.}~\bibnamefont {Candlish}},
  \bibinfo {author} {\bibfnamefont {E.}~\bibnamefont {Badrick}}, \bibinfo
  {author} {\bibfnamefont {A.}~\bibnamefont {Renehan}},\ and\ \bibinfo {author}
  {\bibfnamefont {I.}~\bibnamefont {Buchan}},\ }\bibfield  {title} {\bibinfo
  {title} {Collider bias is only a partial explanation for the obesity
  paradox},\ }\href@noop {} {\bibfield  {journal} {\bibinfo  {journal}
  {Epidemiology (Cambridge, Mass.)}\ }\textbf {\bibinfo {volume} {27}},\
  \bibinfo {pages} {525} (\bibinfo {year} {2016})}\BibitemShut {NoStop}%
\bibitem [{\citenamefont {Coscia}\ \emph {et~al.}(2022)\citenamefont {Coscia},
  \citenamefont {Gill}, \citenamefont {Ben{\'\i}tez}, \citenamefont
  {P{\'e}rez}, \citenamefont {Malats},\ and\ \citenamefont
  {Burgess}}]{coscia2022avoiding}%
  \BibitemOpen
  \bibfield  {author} {\bibinfo {author} {\bibfnamefont {C.}~\bibnamefont
  {Coscia}}, \bibinfo {author} {\bibfnamefont {D.}~\bibnamefont {Gill}},
  \bibinfo {author} {\bibfnamefont {R.}~\bibnamefont {Ben{\'\i}tez}}, \bibinfo
  {author} {\bibfnamefont {T.}~\bibnamefont {P{\'e}rez}}, \bibinfo {author}
  {\bibfnamefont {N.}~\bibnamefont {Malats}},\ and\ \bibinfo {author}
  {\bibfnamefont {S.}~\bibnamefont {Burgess}},\ }\bibfield  {title} {\bibinfo
  {title} {Avoiding collider bias in mendelian randomization when performing
  stratified analyses},\ }\href@noop {} {\bibfield  {journal} {\bibinfo
  {journal} {European Journal of Epidemiology}\ }\textbf {\bibinfo {volume}
  {37}},\ \bibinfo {pages} {671} (\bibinfo {year} {2022})}\BibitemShut
  {NoStop}%
\bibitem [{\citenamefont {Del~Junco}\ \emph {et~al.}(2015)\citenamefont
  {Del~Junco}, \citenamefont {Bulger}, \citenamefont {Fox}, \citenamefont
  {Holcomb}, \citenamefont {Brasel}, \citenamefont {Hoyt}, \citenamefont
  {Grady}, \citenamefont {Duran}, \citenamefont {Klotz}, \citenamefont {Dubick}
  \emph {et~al.}}]{del2015collider}%
  \BibitemOpen
  \bibfield  {author} {\bibinfo {author} {\bibfnamefont {D.~J.}\ \bibnamefont
  {Del~Junco}}, \bibinfo {author} {\bibfnamefont {E.~M.}\ \bibnamefont
  {Bulger}}, \bibinfo {author} {\bibfnamefont {E.~E.}\ \bibnamefont {Fox}},
  \bibinfo {author} {\bibfnamefont {J.~B.}\ \bibnamefont {Holcomb}}, \bibinfo
  {author} {\bibfnamefont {K.~J.}\ \bibnamefont {Brasel}}, \bibinfo {author}
  {\bibfnamefont {D.~B.}\ \bibnamefont {Hoyt}}, \bibinfo {author}
  {\bibfnamefont {J.~J.}\ \bibnamefont {Grady}}, \bibinfo {author}
  {\bibfnamefont {S.}~\bibnamefont {Duran}}, \bibinfo {author} {\bibfnamefont
  {P.}~\bibnamefont {Klotz}}, \bibinfo {author} {\bibfnamefont {M.~A.}\
  \bibnamefont {Dubick}}, \emph {et~al.},\ }\bibfield  {title} {\bibinfo
  {title} {Collider bias in trauma comparative effectiveness research: the
  stratification blues for systematic reviews},\ }\href@noop {} {\bibfield
  {journal} {\bibinfo  {journal} {Injury}\ }\textbf {\bibinfo {volume} {46}},\
  \bibinfo {pages} {775} (\bibinfo {year} {2015})}\BibitemShut {NoStop}%
\bibitem [{\citenamefont {Leite}\ \emph {et~al.}(2020)\citenamefont {Leite},
  \citenamefont {Nascimento}, \citenamefont {Peres}, \citenamefont {Demarco},
  \citenamefont {Horta},\ and\ \citenamefont {Peres}}]{leite2020collider}%
  \BibitemOpen
  \bibfield  {author} {\bibinfo {author} {\bibfnamefont {F.~R.}\ \bibnamefont
  {Leite}}, \bibinfo {author} {\bibfnamefont {G.~G.}\ \bibnamefont
  {Nascimento}}, \bibinfo {author} {\bibfnamefont {K.~G.}\ \bibnamefont
  {Peres}}, \bibinfo {author} {\bibfnamefont {F.~F.}\ \bibnamefont {Demarco}},
  \bibinfo {author} {\bibfnamefont {B.~L.}\ \bibnamefont {Horta}},\ and\
  \bibinfo {author} {\bibfnamefont {M.~A.}\ \bibnamefont {Peres}},\ }\bibfield
  {title} {\bibinfo {title} {Collider bias in the association of periodontitis
  and carotid intima-media thickness},\ }\href@noop {} {\bibfield  {journal}
  {\bibinfo  {journal} {Community Dentistry and Oral Epidemiology}\ }\textbf
  {\bibinfo {volume} {48}},\ \bibinfo {pages} {264} (\bibinfo {year}
  {2020})}\BibitemShut {NoStop}%
\bibitem [{\citenamefont {Sanni~Ali}\ \emph {et~al.}(2013)\citenamefont
  {Sanni~Ali}, \citenamefont {Groenwold}, \citenamefont {Pestman},
  \citenamefont {Belitser}, \citenamefont {Hoes}, \citenamefont {De~Boer},\
  and\ \citenamefont {Klungel}}]{sanni2013time}%
  \BibitemOpen
  \bibfield  {author} {\bibinfo {author} {\bibfnamefont {M.}~\bibnamefont
  {Sanni~Ali}}, \bibinfo {author} {\bibfnamefont {R.~H.}\ \bibnamefont
  {Groenwold}}, \bibinfo {author} {\bibfnamefont {W.~R.}\ \bibnamefont
  {Pestman}}, \bibinfo {author} {\bibfnamefont {S.~V.}\ \bibnamefont
  {Belitser}}, \bibinfo {author} {\bibfnamefont {A.~W.}\ \bibnamefont {Hoes}},
  \bibinfo {author} {\bibfnamefont {A.}~\bibnamefont {De~Boer}},\ and\ \bibinfo
  {author} {\bibfnamefont {O.~H.}\ \bibnamefont {Klungel}},\ }\bibfield
  {title} {\bibinfo {title} {Time-dependent propensity score and
  collider-stratification bias: an example of beta 2-agonist use and the risk
  of coronary heart disease},\ }\href@noop {} {\bibfield  {journal} {\bibinfo
  {journal} {European journal of epidemiology}\ }\textbf {\bibinfo {volume}
  {28}},\ \bibinfo {pages} {291} (\bibinfo {year} {2013})}\BibitemShut
  {NoStop}%
\bibitem [{\citenamefont {T{\"o}nnies}\ \emph {et~al.}(2022)\citenamefont
  {T{\"o}nnies}, \citenamefont {Kahl},\ and\ \citenamefont
  {Kuss}}]{tonnies2022collider}%
  \BibitemOpen
  \bibfield  {author} {\bibinfo {author} {\bibfnamefont {T.}~\bibnamefont
  {T{\"o}nnies}}, \bibinfo {author} {\bibfnamefont {S.}~\bibnamefont {Kahl}},\
  and\ \bibinfo {author} {\bibfnamefont {O.}~\bibnamefont {Kuss}},\ }\bibfield
  {title} {\bibinfo {title} {Collider bias in observational studies.},\
  }\href@noop {} {\bibfield  {journal} {\bibinfo  {journal} {Deutsches
  Aerzteblatt International}\ }\textbf {\bibinfo {volume} {119}} (\bibinfo
  {year} {2022})}\BibitemShut {NoStop}%
\bibitem [{\citenamefont {Holmberg}\ and\ \citenamefont
  {Andersen}(2022)}]{holmberg2022collider}%
  \BibitemOpen
  \bibfield  {author} {\bibinfo {author} {\bibfnamefont {M.~J.}\ \bibnamefont
  {Holmberg}}\ and\ \bibinfo {author} {\bibfnamefont {L.~W.}\ \bibnamefont
  {Andersen}},\ }\bibfield  {title} {\bibinfo {title} {Collider bias},\
  }\href@noop {} {\bibfield  {journal} {\bibinfo  {journal} {Jama}\ }\textbf
  {\bibinfo {volume} {327}},\ \bibinfo {pages} {1282} (\bibinfo {year}
  {2022})}\BibitemShut {NoStop}%
\bibitem [{\citenamefont {Griffith}\ \emph {et~al.}(2020)\citenamefont
  {Griffith}, \citenamefont {Morris}, \citenamefont {Tudball}, \citenamefont
  {Herbert}, \citenamefont {Mancano}, \citenamefont {Pike}, \citenamefont
  {Sharp}, \citenamefont {Sterne}, \citenamefont {Palmer}, \citenamefont
  {Davey~Smith} \emph {et~al.}}]{griffith2020collider}%
  \BibitemOpen
  \bibfield  {author} {\bibinfo {author} {\bibfnamefont {G.~J.}\ \bibnamefont
  {Griffith}}, \bibinfo {author} {\bibfnamefont {T.~T.}\ \bibnamefont
  {Morris}}, \bibinfo {author} {\bibfnamefont {M.~J.}\ \bibnamefont {Tudball}},
  \bibinfo {author} {\bibfnamefont {A.}~\bibnamefont {Herbert}}, \bibinfo
  {author} {\bibfnamefont {G.}~\bibnamefont {Mancano}}, \bibinfo {author}
  {\bibfnamefont {L.}~\bibnamefont {Pike}}, \bibinfo {author} {\bibfnamefont
  {G.~C.}\ \bibnamefont {Sharp}}, \bibinfo {author} {\bibfnamefont
  {J.}~\bibnamefont {Sterne}}, \bibinfo {author} {\bibfnamefont {T.~M.}\
  \bibnamefont {Palmer}}, \bibinfo {author} {\bibfnamefont {G.}~\bibnamefont
  {Davey~Smith}}, \emph {et~al.},\ }\bibfield  {title} {\bibinfo {title}
  {Collider bias undermines our understanding of covid-19 disease risk and
  severity},\ }\href@noop {} {\bibfield  {journal} {\bibinfo  {journal} {Nature
  communications}\ }\textbf {\bibinfo {volume} {11}},\ \bibinfo {pages} {5749}
  (\bibinfo {year} {2020})}\BibitemShut {NoStop}%
\bibitem [{\citenamefont {Hern{\'a}n}\ \emph {et~al.}(2004)\citenamefont
  {Hern{\'a}n}, \citenamefont {Hern{\'a}ndez-D{\'\i}az},\ and\ \citenamefont
  {Robins}}]{hernan2004structural}%
  \BibitemOpen
  \bibfield  {author} {\bibinfo {author} {\bibfnamefont {M.~A.}\ \bibnamefont
  {Hern{\'a}n}}, \bibinfo {author} {\bibfnamefont {S.}~\bibnamefont
  {Hern{\'a}ndez-D{\'\i}az}},\ and\ \bibinfo {author} {\bibfnamefont {J.~M.}\
  \bibnamefont {Robins}},\ }\bibfield  {title} {\bibinfo {title} {A structural
  approach to selection bias},\ }\href@noop {} {\bibfield  {journal} {\bibinfo
  {journal} {Epidemiology}\ ,\ \bibinfo {pages} {615}} (\bibinfo {year}
  {2004})}\BibitemShut {NoStop}%
\bibitem [{\citenamefont {VanderWeele}\ and\ \citenamefont
  {Robins}(2007)}]{vanderweele2007directed}%
  \BibitemOpen
  \bibfield  {author} {\bibinfo {author} {\bibfnamefont {T.~J.}\ \bibnamefont
  {VanderWeele}}\ and\ \bibinfo {author} {\bibfnamefont {J.~M.}\ \bibnamefont
  {Robins}},\ }\bibfield  {title} {\bibinfo {title} {Directed acyclic graphs,
  sufficient causes, and the properties of conditioning on a common effect},\
  }\href@noop {} {\bibfield  {journal} {\bibinfo  {journal} {American journal
  of epidemiology}\ }\textbf {\bibinfo {volume} {166}},\ \bibinfo {pages}
  {1096} (\bibinfo {year} {2007})}\BibitemShut {NoStop}%
\bibitem [{\citenamefont {Cole}\ \emph {et~al.}(2010)\citenamefont {Cole},
  \citenamefont {Platt}, \citenamefont {Schisterman}, \citenamefont {Chu},
  \citenamefont {Westreich}, \citenamefont {Richardson},\ and\ \citenamefont
  {Poole}}]{cole2010illustrating}%
  \BibitemOpen
  \bibfield  {author} {\bibinfo {author} {\bibfnamefont {S.~R.}\ \bibnamefont
  {Cole}}, \bibinfo {author} {\bibfnamefont {R.~W.}\ \bibnamefont {Platt}},
  \bibinfo {author} {\bibfnamefont {E.~F.}\ \bibnamefont {Schisterman}},
  \bibinfo {author} {\bibfnamefont {H.}~\bibnamefont {Chu}}, \bibinfo {author}
  {\bibfnamefont {D.}~\bibnamefont {Westreich}}, \bibinfo {author}
  {\bibfnamefont {D.}~\bibnamefont {Richardson}},\ and\ \bibinfo {author}
  {\bibfnamefont {C.}~\bibnamefont {Poole}},\ }\bibfield  {title} {\bibinfo
  {title} {Illustrating bias due to conditioning on a collider},\ }\href@noop
  {} {\bibfield  {journal} {\bibinfo  {journal} {International journal of
  epidemiology}\ }\textbf {\bibinfo {volume} {39}},\ \bibinfo {pages} {417}
  (\bibinfo {year} {2010})}\BibitemShut {NoStop}%
\bibitem [{\citenamefont {Weissman}(2020)}]{weissman2020gre}%
  \BibitemOpen
  \bibfield  {author} {\bibinfo {author} {\bibfnamefont {M.~B.}\ \bibnamefont
  {Weissman}},\ }\bibfield  {title} {\bibinfo {title} {Do gre scores help
  predict getting a physics ph. d.? a comment on a paper by miller et al.},\
  }\href@noop {} {\bibfield  {journal} {\bibinfo  {journal} {Science Advances}\
  }\textbf {\bibinfo {volume} {6}} (\bibinfo {year} {2020})}\BibitemShut
  {NoStop}%
\bibitem [{\citenamefont {Yerushalmy}(1971)}]{Yerushalmy1971}%
  \BibitemOpen
  \bibfield  {author} {\bibinfo {author} {\bibfnamefont {J.}~\bibnamefont
  {Yerushalmy}},\ }\bibfield  {title} {\bibinfo {title} {The relationship of
  parents' cigarette smoking to outcome of pregnancy--implications as to the
  problem of inferring causation from observed associations.},\ }\href@noop {}
  {\bibfield  {journal} {\bibinfo  {journal} {American journal of
  epidemiology}\ }\textbf {\bibinfo {volume} {93(6)}},\ \bibinfo {pages} {443}
  (\bibinfo {year} {1971})}\BibitemShut {NoStop}%
\bibitem [{\citenamefont {Nissen}\ \emph {et~al.}(2018)\citenamefont {Nissen},
  \citenamefont {Jariwala}, \citenamefont {Close},\ and\ \citenamefont
  {Dusen}}]{nissen2018participation}%
  \BibitemOpen
  \bibfield  {author} {\bibinfo {author} {\bibfnamefont {J.~M.}\ \bibnamefont
  {Nissen}}, \bibinfo {author} {\bibfnamefont {M.}~\bibnamefont {Jariwala}},
  \bibinfo {author} {\bibfnamefont {E.~W.}\ \bibnamefont {Close}},\ and\
  \bibinfo {author} {\bibfnamefont {B.~V.}\ \bibnamefont {Dusen}},\ }\bibfield
  {title} {\bibinfo {title} {Participation and performance on paper-and
  computer-based low-stakes assessments},\ }\href@noop {} {\bibfield  {journal}
  {\bibinfo  {journal} {International journal of STEM education}\ }\textbf
  {\bibinfo {volume} {5}},\ \bibinfo {pages} {1} (\bibinfo {year}
  {2018})}\BibitemShut {NoStop}%
\bibitem [{\citenamefont {Nissen}\ \emph {et~al.}(2019)\citenamefont {Nissen},
  \citenamefont {Donatello},\ and\ \citenamefont
  {Van~Dusen}}]{nissen2019missing}%
  \BibitemOpen
  \bibfield  {author} {\bibinfo {author} {\bibfnamefont {J.}~\bibnamefont
  {Nissen}}, \bibinfo {author} {\bibfnamefont {R.}~\bibnamefont {Donatello}},\
  and\ \bibinfo {author} {\bibfnamefont {B.}~\bibnamefont {Van~Dusen}},\
  }\bibfield  {title} {\bibinfo {title} {Missing data and bias in physics
  education research: A case for using multiple imputation},\ }\href@noop {}
  {\bibfield  {journal} {\bibinfo  {journal} {Physical Review Physics Education
  Research}\ }\textbf {\bibinfo {volume} {15}},\ \bibinfo {pages} {020106}
  (\bibinfo {year} {2019})}\BibitemShut {NoStop}%
\bibitem [{\citenamefont {Bandura}(1986)}]{bandura1986social}%
  \BibitemOpen
  \bibfield  {author} {\bibinfo {author} {\bibfnamefont {A.}~\bibnamefont
  {Bandura}},\ }\bibfield  {title} {\bibinfo {title} {Social foundations of
  thought and action},\ }\href@noop {} {\bibfield  {journal} {\bibinfo
  {journal} {Englewood Cliffs, NJ}\ }\textbf {\bibinfo {volume} {1986}},\
  \bibinfo {pages} {23} (\bibinfo {year} {1986})}\BibitemShut {NoStop}%
\bibitem [{\citenamefont {Locke}(1997)}]{locke1997self}%
  \BibitemOpen
  \bibfield  {author} {\bibinfo {author} {\bibfnamefont {E.~A.}\ \bibnamefont
  {Locke}},\ }\bibfield  {title} {\bibinfo {title} {Self-efficacy: The exercise
  of control},\ }\href@noop {} {\bibfield  {journal} {\bibinfo  {journal}
  {Personnel psychology}\ }\textbf {\bibinfo {volume} {50}},\ \bibinfo {pages}
  {801} (\bibinfo {year} {1997})}\BibitemShut {NoStop}%
\bibitem [{\citenamefont {Liem}\ \emph {et~al.}(2008)\citenamefont {Liem},
  \citenamefont {Lau},\ and\ \citenamefont {Nie}}]{liem2008role}%
  \BibitemOpen
  \bibfield  {author} {\bibinfo {author} {\bibfnamefont {A.~D.}\ \bibnamefont
  {Liem}}, \bibinfo {author} {\bibfnamefont {S.}~\bibnamefont {Lau}},\ and\
  \bibinfo {author} {\bibfnamefont {Y.}~\bibnamefont {Nie}},\ }\bibfield
  {title} {\bibinfo {title} {The role of self-efficacy, task value, and
  achievement goals in predicting learning strategies, task disengagement, peer
  relationship, and achievement outcome},\ }\href@noop {} {\bibfield  {journal}
  {\bibinfo  {journal} {Contemporary educational psychology}\ }\textbf
  {\bibinfo {volume} {33}},\ \bibinfo {pages} {486} (\bibinfo {year}
  {2008})}\BibitemShut {NoStop}%
\bibitem [{\citenamefont {Vogt}(2008)}]{vogt2008faculty}%
  \BibitemOpen
  \bibfield  {author} {\bibinfo {author} {\bibfnamefont {C.~M.}\ \bibnamefont
  {Vogt}},\ }\bibfield  {title} {\bibinfo {title} {Faculty as a critical
  juncture in student retention and performance in engineering programs},\
  }\href@noop {} {\bibfield  {journal} {\bibinfo  {journal} {Journal of
  Engineering Education}\ }\textbf {\bibinfo {volume} {97}},\ \bibinfo {pages}
  {27} (\bibinfo {year} {2008})}\BibitemShut {NoStop}%
\bibitem [{\citenamefont {Jones}\ \emph {et~al.}(2010)\citenamefont {Jones},
  \citenamefont {Paretti}, \citenamefont {Hein},\ and\ \citenamefont
  {Knott}}]{jones2010analysis}%
  \BibitemOpen
  \bibfield  {author} {\bibinfo {author} {\bibfnamefont {B.~D.}\ \bibnamefont
  {Jones}}, \bibinfo {author} {\bibfnamefont {M.~C.}\ \bibnamefont {Paretti}},
  \bibinfo {author} {\bibfnamefont {S.~F.}\ \bibnamefont {Hein}},\ and\
  \bibinfo {author} {\bibfnamefont {T.~W.}\ \bibnamefont {Knott}},\ }\bibfield
  {title} {\bibinfo {title} {An analysis of motivation constructs with
  first-year engineering students: Relationships among expectancies, values,
  achievement, and career plans},\ }\href@noop {} {\bibfield  {journal}
  {\bibinfo  {journal} {Journal of engineering education}\ }\textbf {\bibinfo
  {volume} {99}},\ \bibinfo {pages} {319} (\bibinfo {year} {2010})}\BibitemShut
  {NoStop}%
\bibitem [{\citenamefont {Honicke}\ and\ \citenamefont
  {Broadbent}(2016)}]{honicke2016influence}%
  \BibitemOpen
  \bibfield  {author} {\bibinfo {author} {\bibfnamefont {T.}~\bibnamefont
  {Honicke}}\ and\ \bibinfo {author} {\bibfnamefont {J.}~\bibnamefont
  {Broadbent}},\ }\bibfield  {title} {\bibinfo {title} {The influence of
  academic self-efficacy on academic performance: A systematic review},\
  }\href@noop {} {\bibfield  {journal} {\bibinfo  {journal} {Educational
  research review}\ }\textbf {\bibinfo {volume} {17}},\ \bibinfo {pages} {63}
  (\bibinfo {year} {2016})}\BibitemShut {NoStop}%
\bibitem [{\citenamefont {Vancouver}\ \emph {et~al.}(2001)\citenamefont
  {Vancouver}, \citenamefont {Thompson},\ and\ \citenamefont
  {Williams}}]{vancouver2001changing}%
  \BibitemOpen
  \bibfield  {author} {\bibinfo {author} {\bibfnamefont {J.~B.}\ \bibnamefont
  {Vancouver}}, \bibinfo {author} {\bibfnamefont {C.~M.}\ \bibnamefont
  {Thompson}},\ and\ \bibinfo {author} {\bibfnamefont {A.~A.}\ \bibnamefont
  {Williams}},\ }\bibfield  {title} {\bibinfo {title} {The changing signs in
  the relationships among self-efficacy, personal goals, and performance.},\
  }\href@noop {} {\bibfield  {journal} {\bibinfo  {journal} {Journal of applied
  psychology}\ }\textbf {\bibinfo {volume} {86}},\ \bibinfo {pages} {605}
  (\bibinfo {year} {2001})}\BibitemShut {NoStop}%
\bibitem [{\citenamefont {Hattie}\ and\ \citenamefont
  {Anderman}(2013)}]{hattie2013international}%
  \BibitemOpen
  \bibfield  {author} {\bibinfo {author} {\bibfnamefont {J.}~\bibnamefont
  {Hattie}}\ and\ \bibinfo {author} {\bibfnamefont {E.~M.}\ \bibnamefont
  {Anderman}},\ }\href@noop {} {\emph {\bibinfo {title} {International guide to
  student achievement}}}\ (\bibinfo  {publisher} {Routledge},\ \bibinfo {year}
  {2013})\BibitemShut {NoStop}%
\bibitem [{\citenamefont {Pajares}(1997)}]{pajares1997current}%
  \BibitemOpen
  \bibfield  {author} {\bibinfo {author} {\bibfnamefont {F.}~\bibnamefont
  {Pajares}},\ }\bibfield  {title} {\bibinfo {title} {Current directions in
  self-efficacy research},\ }\href@noop {} {\bibfield  {journal} {\bibinfo
  {journal} {Advances in motivation and achievement}\ }\textbf {\bibinfo
  {volume} {10}},\ \bibinfo {pages} {1} (\bibinfo {year} {1997})}\BibitemShut
  {NoStop}%
\bibitem [{\citenamefont {Zimmerman}(2000)}]{zimmerman2000self}%
  \BibitemOpen
  \bibfield  {author} {\bibinfo {author} {\bibfnamefont {B.~J.}\ \bibnamefont
  {Zimmerman}},\ }\bibfield  {title} {\bibinfo {title} {Self-efficacy: An
  essential motive to learn},\ }\href@noop {} {\bibfield  {journal} {\bibinfo
  {journal} {Contemporary educational psychology}\ }\textbf {\bibinfo {volume}
  {25}},\ \bibinfo {pages} {82} (\bibinfo {year} {2000})}\BibitemShut {NoStop}%
\bibitem [{\citenamefont {Britner}\ and\ \citenamefont
  {Pajares}(2006)}]{britner2006sources}%
  \BibitemOpen
  \bibfield  {author} {\bibinfo {author} {\bibfnamefont {S.~L.}\ \bibnamefont
  {Britner}}\ and\ \bibinfo {author} {\bibfnamefont {F.}~\bibnamefont
  {Pajares}},\ }\bibfield  {title} {\bibinfo {title} {Sources of science
  self-efficacy beliefs of middle school students},\ }\href@noop {} {\bibfield
  {journal} {\bibinfo  {journal} {Journal of Research in Science Teaching: The
  Official Journal of the National Association for Research in Science
  Teaching}\ }\textbf {\bibinfo {volume} {43}},\ \bibinfo {pages} {485}
  (\bibinfo {year} {2006})}\BibitemShut {NoStop}%
\bibitem [{\citenamefont {Usher}\ and\ \citenamefont
  {Pajares}(2008)}]{usher2008sources}%
  \BibitemOpen
  \bibfield  {author} {\bibinfo {author} {\bibfnamefont {E.~L.}\ \bibnamefont
  {Usher}}\ and\ \bibinfo {author} {\bibfnamefont {F.}~\bibnamefont
  {Pajares}},\ }\bibfield  {title} {\bibinfo {title} {Sources of self-efficacy
  in school: Critical review of the literature and future directions},\
  }\href@noop {} {\bibfield  {journal} {\bibinfo  {journal} {Review of
  educational research}\ }\textbf {\bibinfo {volume} {78}},\ \bibinfo {pages}
  {751} (\bibinfo {year} {2008})}\BibitemShut {NoStop}%
\bibitem [{\citenamefont {Lent}\ \emph {et~al.}(1991)\citenamefont {Lent},
  \citenamefont {Lopez},\ and\ \citenamefont {Bieschke}}]{lent1991mathematics}%
  \BibitemOpen
  \bibfield  {author} {\bibinfo {author} {\bibfnamefont {R.~W.}\ \bibnamefont
  {Lent}}, \bibinfo {author} {\bibfnamefont {F.~G.}\ \bibnamefont {Lopez}},\
  and\ \bibinfo {author} {\bibfnamefont {K.~J.}\ \bibnamefont {Bieschke}},\
  }\bibfield  {title} {\bibinfo {title} {Mathematics self-efficacy: Sources and
  relation to science-based career choice.},\ }\href@noop {} {\bibfield
  {journal} {\bibinfo  {journal} {Journal of counseling psychology}\ }\textbf
  {\bibinfo {volume} {38}},\ \bibinfo {pages} {424} (\bibinfo {year}
  {1991})}\BibitemShut {NoStop}%
\bibitem [{\citenamefont {Klassen}(2004)}]{klassen2004cross}%
  \BibitemOpen
  \bibfield  {author} {\bibinfo {author} {\bibfnamefont {R.~M.}\ \bibnamefont
  {Klassen}},\ }\bibfield  {title} {\bibinfo {title} {A cross-cultural
  investigation of the efficacy beliefs of south asian immigrant and anglo
  canadian nonimmigrant early adolescents.},\ }\href@noop {} {\bibfield
  {journal} {\bibinfo  {journal} {Journal of Educational Psychology}\ }\textbf
  {\bibinfo {volume} {96}},\ \bibinfo {pages} {731} (\bibinfo {year}
  {2004})}\BibitemShut {NoStop}%
\bibitem [{\citenamefont {Matsui}\ \emph {et~al.}(1990)\citenamefont {Matsui},
  \citenamefont {Matsui},\ and\ \citenamefont
  {Ohnishi}}]{matsui1990mechanisms}%
  \BibitemOpen
  \bibfield  {author} {\bibinfo {author} {\bibfnamefont {T.}~\bibnamefont
  {Matsui}}, \bibinfo {author} {\bibfnamefont {K.}~\bibnamefont {Matsui}},\
  and\ \bibinfo {author} {\bibfnamefont {R.}~\bibnamefont {Ohnishi}},\
  }\bibfield  {title} {\bibinfo {title} {Mechanisms underlying math
  self-efficacy learning of college students},\ }\href@noop {} {\bibfield
  {journal} {\bibinfo  {journal} {Journal of Vocational Behavior}\ }\textbf
  {\bibinfo {volume} {37}},\ \bibinfo {pages} {225} (\bibinfo {year}
  {1990})}\BibitemShut {NoStop}%
\bibitem [{\citenamefont {Multon}\ \emph {et~al.}(1991)\citenamefont {Multon},
  \citenamefont {Brown},\ and\ \citenamefont {Lent}}]{multon1991relation}%
  \BibitemOpen
  \bibfield  {author} {\bibinfo {author} {\bibfnamefont {K.~D.}\ \bibnamefont
  {Multon}}, \bibinfo {author} {\bibfnamefont {S.~D.}\ \bibnamefont {Brown}},\
  and\ \bibinfo {author} {\bibfnamefont {R.~W.}\ \bibnamefont {Lent}},\
  }\bibfield  {title} {\bibinfo {title} {Relation of self-efficacy beliefs to
  academic outcomes: A meta-analytic investigation.},\ }\href@noop {}
  {\bibfield  {journal} {\bibinfo  {journal} {Journal of counseling
  psychology}\ }\textbf {\bibinfo {volume} {38}},\ \bibinfo {pages} {30}
  (\bibinfo {year} {1991})}\BibitemShut {NoStop}%
\bibitem [{\citenamefont {Phan}(2012)}]{phan2012informational}%
  \BibitemOpen
  \bibfield  {author} {\bibinfo {author} {\bibfnamefont {H.~P.}\ \bibnamefont
  {Phan}},\ }\bibfield  {title} {\bibinfo {title} {Informational sources,
  self-efficacy and achievement: A temporally displaced approach},\ }\href@noop
  {} {\bibfield  {journal} {\bibinfo  {journal} {Educational Psychology}\
  }\textbf {\bibinfo {volume} {32}},\ \bibinfo {pages} {699} (\bibinfo {year}
  {2012})}\BibitemShut {NoStop}%
\bibitem [{\citenamefont {Boden}\ \emph {et~al.}(2018)\citenamefont {Boden},
  \citenamefont {Kuo}, \citenamefont {Nokes-Malach}, \citenamefont {Wallace},\
  and\ \citenamefont {Menekse}}]{boden2018role}%
  \BibitemOpen
  \bibfield  {author} {\bibinfo {author} {\bibfnamefont {K.}~\bibnamefont
  {Boden}}, \bibinfo {author} {\bibfnamefont {E.}~\bibnamefont {Kuo}}, \bibinfo
  {author} {\bibfnamefont {T.}~\bibnamefont {Nokes-Malach}}, \bibinfo {author}
  {\bibfnamefont {T.}~\bibnamefont {Wallace}},\ and\ \bibinfo {author}
  {\bibfnamefont {M.}~\bibnamefont {Menekse}},\ }\bibfield  {title} {\bibinfo
  {title} {What is the role of motivation in procedural and conceptual physics
  learning? an examination of self-efficacy and achievement goals},\ }in\
  \href@noop {} {\emph {\bibinfo {booktitle} {Proceedings of the 2017 Physics
  Education Research Conference, Cincinnati, OH}}}\ (\bibinfo {year} {2018})\
  p.~\bibinfo {pages} {60}\BibitemShut {NoStop}%
\bibitem [{\citenamefont {Talsma}\ \emph {et~al.}(2018)\citenamefont {Talsma},
  \citenamefont {Sch{\"u}z}, \citenamefont {Schwarzer},\ and\ \citenamefont
  {Norris}}]{talsma2018believe}%
  \BibitemOpen
  \bibfield  {author} {\bibinfo {author} {\bibfnamefont {K.}~\bibnamefont
  {Talsma}}, \bibinfo {author} {\bibfnamefont {B.}~\bibnamefont {Sch{\"u}z}},
  \bibinfo {author} {\bibfnamefont {R.}~\bibnamefont {Schwarzer}},\ and\
  \bibinfo {author} {\bibfnamefont {K.}~\bibnamefont {Norris}},\ }\bibfield
  {title} {\bibinfo {title} {I believe, therefore i achieve (and vice versa): A
  meta-analytic cross-lagged panel analysis of self-efficacy and academic
  performance},\ }\href@noop {} {\bibfield  {journal} {\bibinfo  {journal}
  {Learning and Individual Differences}\ }\textbf {\bibinfo {volume} {61}},\
  \bibinfo {pages} {136} (\bibinfo {year} {2018})}\BibitemShut {NoStop}%
\bibitem [{\citenamefont {Mund}\ and\ \citenamefont
  {Nestler}(2019)}]{mund2019beyond}%
  \BibitemOpen
  \bibfield  {author} {\bibinfo {author} {\bibfnamefont {M.}~\bibnamefont
  {Mund}}\ and\ \bibinfo {author} {\bibfnamefont {S.}~\bibnamefont {Nestler}},\
  }\bibfield  {title} {\bibinfo {title} {Beyond the cross-lagged panel model:
  Next-generation statistical tools for analyzing interdependencies across the
  life course},\ }\href@noop {} {\bibfield  {journal} {\bibinfo  {journal}
  {Advances in Life Course Research}\ }\textbf {\bibinfo {volume} {41}},\
  \bibinfo {pages} {100249} (\bibinfo {year} {2019})}\BibitemShut {NoStop}%
\bibitem [{\citenamefont {Hamaker}\ \emph {et~al.}(2015)\citenamefont
  {Hamaker}, \citenamefont {Kuiper},\ and\ \citenamefont
  {Grasman}}]{hamaker2015critique}%
  \BibitemOpen
  \bibfield  {author} {\bibinfo {author} {\bibfnamefont {E.~L.}\ \bibnamefont
  {Hamaker}}, \bibinfo {author} {\bibfnamefont {R.~M.}\ \bibnamefont
  {Kuiper}},\ and\ \bibinfo {author} {\bibfnamefont {R.~P.}\ \bibnamefont
  {Grasman}},\ }\bibfield  {title} {\bibinfo {title} {A critique of the
  cross-lagged panel model.},\ }\href@noop {} {\bibfield  {journal} {\bibinfo
  {journal} {Psychological methods}\ }\textbf {\bibinfo {volume} {20}},\
  \bibinfo {pages} {102} (\bibinfo {year} {2015})}\BibitemShut {NoStop}%
\bibitem [{\citenamefont {Usami}\ \emph {et~al.}(2019)\citenamefont {Usami},
  \citenamefont {Todo},\ and\ \citenamefont {Murayama}}]{usami2019modeling}%
  \BibitemOpen
  \bibfield  {author} {\bibinfo {author} {\bibfnamefont {S.}~\bibnamefont
  {Usami}}, \bibinfo {author} {\bibfnamefont {N.}~\bibnamefont {Todo}},\ and\
  \bibinfo {author} {\bibfnamefont {K.}~\bibnamefont {Murayama}},\ }\bibfield
  {title} {\bibinfo {title} {Modeling reciprocal effects in medical research:
  Critical discussion on the current practices and potential alternative
  models},\ }\href@noop {} {\bibfield  {journal} {\bibinfo  {journal} {PloS
  one}\ }\textbf {\bibinfo {volume} {14}},\ \bibinfo {pages} {e0209133}
  (\bibinfo {year} {2019})}\BibitemShut {NoStop}%
\bibitem [{\citenamefont {Li}\ and\ \citenamefont
  {Singh}(2023)}]{li2023select}%
  \BibitemOpen
  \bibfield  {author} {\bibinfo {author} {\bibfnamefont {Y.}~\bibnamefont
  {Li}}\ and\ \bibinfo {author} {\bibfnamefont {C.}~\bibnamefont {Singh}},\
  }\bibfield  {title} {\bibinfo {title} {How to select suitable models from
  many statistically equivalent models: An example from physics identity},\
  }\href@noop {} {\bibfield  {journal} {\bibinfo  {journal} {arXiv preprint
  arXiv:2303.13786}\ } (\bibinfo {year} {2023})}\BibitemShut {NoStop}%
\end{thebibliography}%

\end{document}